\documentclass[acmsmall,nonacm.anonymous]{acmart}

\acmSubmissionID{110}
\renewcommand\footnotetextcopyrightpermission[1]{}

\usepackage{algorithmic}
\usepackage{graphicx}
\usepackage{textcomp}
\usepackage{xspace}
\usepackage{xcolor}
\usepackage{comment}
\usepackage{grumble}
\usepackage{xurl}
\usepackage{listings}

\usepackage{amsmath}

\newcommand{\hc}{\ensuremath{h_c}}
\newcommand{\sigmac}{\ensuremath{\sigma_c}}
\newcommand{\hr}{\ensuremath{h_r}}
\newcommand{\sigmacq}{\ensuremath{\sigma_{c_q}}}
\newcommand{\kv}{\ensuremath{kv}}
\newcommand{\alphaa}{\ensuremath{\alpha}}
\newcommand{\cq}{\ensuremath{c_q}}
\newcommand{\cntcq}{\ensuremath{cnt_{c_q}}}

\newcommand{\view}{\ensuremath{view}}
\newcommand{\at}{\ensuremath{\,@\,}}

\usepackage{balance}
\usepackage{subcaption}
\usepackage{libertine}
\usepackage{enumitem}
\usepackage{colortbl}
\usepackage{pifont}
\usepackage{tikz}
\usepackage{tcolorbox}

\usepackage{multirow}
\usepackage{algorithmic}
\usepackage[binary-units=true]{siunitx}
\usepackage[ruled,vlined,linesnumbered]{algorithm2e}
\newcommand*\circled[1]{\tikz[baseline=(char.base)]{
            \node[shape=circle,draw,inner sep=0.5pt] (char) {\scriptsize#1};}}
\usepackage{array}

\SetKwComment{Comment}{$\triangleright$ }{}%
\def\BibTeX{{\rm B\kern-.05em{\sc i\kern-.025em b}\kern-.08em
    T\kern-.1667em\lower.7ex\hbox{E}\kern-.125emX}}

\newcommand{\myparagraph}[1]{\smallskip \noindent{\bf {#1}.}}

\newcommand{\out}[1] {}

\newcounter{codeLineCntr}

\reversemarginpar
\newif\ifnotes
\notestrue

\newcommand{\punt}[1]{}

\renewcommand{\eqref}[1]{Equation~(\ref{eq:#1})}

\newcommand{\proc}[1]{\ifmmode\mbox{\textsc{#1}}\else\textsc{#1}\fi}

  \newcommand{\func}[1]{\ifmmode\mathrm{#1}\else\textrm{#1}fi} %

\newcounter{remark}[section]

\usepackage[english]{babel}
\newcommand{\projecttitle}{\textsc{Recipe}\xspace}
\newcommand{\scone}{\textsc{Scone}\xspace}
\definecolor{codegreen}{rgb}{0,0.6,0}
\definecolor{codegray}{rgb}{0.5,0.5,0.5}
\definecolor{codepurple}{rgb}{0.58,0,0.82}
\definecolor{backcolour}{rgb}{0.95,0.95,0.92}

\definecolor{lightGrey}{rgb}{0.9, 0.9, 0.9}
\definecolor{beaublue}{rgb}{0.74, 0.83, 0.9}
\definecolor{lightred}{RGB}{229, 220, 220}

\definecolor{burlywood}{rgb}{0.87, 0.72, 0.53}

\lstdefinestyle{customc}{
    backgroundcolor=\color{backcolour},   
    commentstyle=\color{codegreen},
    keywordstyle=\color{magenta},
    numberstyle=\tiny\color{codegray},
    stringstyle=\color{codepurple},
    basicstyle=\ttfamily\footnotesize,
    breaklines,
    tabsize=2,
    numbers=left,
    columns=fullflexible,
    keepspaces=true,
    frame=lines,
    numbersep=4pt,
    escapechar=@,
    mathescape=true,
    captionpos=b,
    language=c++,
    keywords = {auto, new, void, Raft_ctx, Msg, for}
}

\renewcommand{\subparagraph}{}

\usepackage[small,compact,noindentafter]{titlesec} 
\usepackage[compact]{titlesec}

\titlespacing\section{0pt}{6pt plus 2pt minus 2pt}{2pt plus 2pt minus 2pt}
\titlespacing\subsection{0pt}{4pt plus 2pt minus 2pt}{2pt plus 1pt minus 1pt}
\titlespacing{\paragraph}{0pt}{2pt plus 0pt minus 1pt}{1.0ex}

\usepackage[subtle,margins=normal]{savetrees}

\usepackage{setspace}
\usepackage[margin=5pt,font={stretch=0.9}]{caption}

\usepackage{amssymb}

\begin{document}
\title{\projecttitle: Hardware-Accelerated Replication Protocols}
\subtitle{Rethinking Crash Fault Tolerance Protocols for Modern Cloud Environments}
\author{Dimitra Giantsidi}
\affiliation{%
\institution{The University of Edinburgh}
\country{UK}
}

\author{Emmanouil Giortamis}
\affiliation{%
\institution{TU Munich}
\country{Germany}
}

\author{Julian Pritzi}
\affiliation{%
\institution{TU Munich}
\country{Germany}
}

\author{Maurice Bailleu}
\affiliation{%
\institution{Huawei Research Edinburgh}
\country{UK}
}

\author{Manos Kapritsos}
\affiliation{%
\institution{University of Michigan}
\country{USA}
}

\author{Pramod Bhatotia}
\affiliation{%
\institution{TU Munich}
\country{Germany}
}

\begin{abstract}

Replication protocols are fundamental to distributed systems, ensuring consistency, reliability, and fault tolerance. Traditional Crash Fault Tolerant (CFT) replication protocols assume a fail-stop model, making them unsuitable for untrusted cloud environments where adversaries (e.g., co-located tenants or a malicious administrator) or software bugs may lead to Byzantine behavior. Byzantine Fault Tolerant (BFT) protocols address these threats but suffer significant performance, resource overheads, and programmability and scalability challenges.

This paper presents \projecttitle{}, a novel approach for transforming CFT protocols to operate securely in Byzantine settings without modifying their core logic. Our approach is to rethink the existing CFT protocols in today’s modern hardware in cloud environments, with many core servers, RDMA-capable networks, and trusted execution environments. Modern hardware challenges the conventional wisdom on CFT protocol design; our work explores the synergy between modern hardware and the security and performance of strongly consistent replication protocols. \projecttitle{} leverages these advances to rethink CFT protocols for untrusted cloud environments. Specifically, we ask the following question:  \emph{can we leverage (and how) modern cloud hardware to harden the security properties of a CFT protocol for Byzantine settings while achieving high performance?}

\projecttitle{} leverages Trusted Execution Environments (TEEs) and high-performance networking stacks (e.g., RDMA, DPDK) to implement two practical security mechanisms that ensure the transferable authentication and non-equivocation properties. These two properties are the lower bound for transforming any CFT protocol for Byzantine settings, guaranteeing that a transformation of (any) CFT protocol to a BFT one \emph{always} exists.

\projecttitle{} protocol integrates modern hardware (TEEs and high-performant network stacks) to design five key components: a Transferable authentication phase that ensures the authenticity of all participants, the Initialization phase, the Normal operation phase, which executes clients' requests as well as the View Change and Recovery. \projecttitle{} protocol is verified formally with Tamarin, a symbolic model checker confirming our transformation's correctness. We implemented \projecttitle{} protocol as a library, and we applied to transform four widely used CFT protocols to operate on Byzantine settings—Raft, Chain Replication, ABD, and AllConcur—demonstrating up to 24× higher throughput compared to PBFT and 5.9× better performance than state-of-the-art BFT protocols. \projecttitle{} achieves these gains while requiring $f$ fewer replicas and offering confidentiality, a feature absent in traditional BFT protocols.

\end{abstract}

\maketitle



\section{Introduction}
\label{sec:introduction}

Replication protocols play a foundational role in designing distributed systems, such as distributed data storage systems~\cite{10.14778/3007263.3007267, 10.14778/2002938.2002939, zippydb, bankDB, NetflixDB, GEDB, HESSDB}, distributed coordination services~\cite{Hunt:2010, chubby},  distributed ledgers~\cite{baasAlibaba, baasAWS, baasAzure, baasIBM, baasOracle}, and distributed data analytics systems~\cite{NetflixDB}. For performance and fault tolerance requirements, distributed systems employ Crash Fault Tolerant (CFT) {\em replication protocols}~\cite{raft,chain-replication, 10.5555/1855807.1855818, Reed2008AST, 10.5555/800253.807732, lynch:1997, 10.1145/279227.279229, Hermes:2020} to maintain a consistent view of the datasets, guaranteeing fault tolerance, i.e., reliability and availability in the presence of failures~\cite{zippydb, dynamo, lakshman2009, redis, rocksdb, leveldb, memcached2004, bfthyperledger}.

Unfortunately, CFT protocols assume a {\em fail-stop model}, i.e., replicas are honest and can only fail by crashing~\cite{delporte}. As such, they are {\em inadequate} for modern untrusted cloud environments, where the underlying cloud infrastructure can be compromised by an adversary, e.g., co-located tenants or even a misbehaving cloud operator that may eavesdrop or actively influence the replicas' behavior.
In such an untrusted environment, the surface of faults and attacks expands beyond the CFT fail-stop model, ranging from software bugs and configuration errors to malicious attacks~\cite{Gunawi_bugs-in-the-cloud, Shinde2016, hahnel2017high}. CFT protocols are fundamentally incapable of providing consistent replication in the presence of non-benign ({\em Byzantine}) faults in untrusted cloud environments.

\subsection{The CFT Vs. BFT Conundrum} 

\myparagraph{CFT protocols}
CFT protocols assume that the infrastructure is trusted. These protocols tolerate only benign faults; replicas can fail by stopping or by omitting some steps~\cite{delporte}. As such, while having low overheads, they are not suitable for modern applications deployed in third-party untrusted cloud infrastructure~\cite{10.1145/3190508.3190538}. In this paper, we evaluate protocols that enforce either sequential consistency~\cite{1675439} or linearizability~\cite{Herlihy:1990}, also referred to as \emph{strongly-consistent} replication protocols.

We can broadly split strongly-consistent CFT protocols into two categories (see Table~\ref{tab:categories} for the taxonomy):  \emph{(i)} leader-based protocols (e.g., Raft~\cite{raft}, Chain Replication (CR)~\cite{chain-replication}), where a node, designated as a leader, drives the protocol execution and \emph{(ii)} decentralized protocols (e.g., ABD~\cite{lynch:1997},  AllConcur~\cite{Poke2016AllConcurLC}), where there is no leader and all nodes can propose and execute requests. 

We further divide them based on their ordering semantics. First, protocols with total ordering, where the protocols create a total order of all writes across all keys and apply them in that order. Second, protocols with per-key ordering semantics where the protocol enforces the total order of writes on a per-key basis. The evaluation of \projecttitle{} ($\S$~\ref{sec:eval}) relies on this taxonomy to systematically study its protocols' performance, as these two dimensions significantly impact the performance of the CFT protocols~\cite{f04eb9b864204bab958e72055062748c}. 



\begin{table}[t]
\small

\fontsize{7}{10}\selectfont 
\begin{center}
\begin{tabular}{ | p{2.8cm} | p{3.6cm} | p{3.6cm} | }
\rowcolor{gray!25}
 \hline
 & \bf{Leader-based} & \bf{Leader-less} \\ \hline
     \multirow{2}{*}{\bf{Total order}}  & {\bf{Raft}}~\cite{raft}, ZAB~\cite{Reed2008AST},            & {\bf{AllConcur}}~\cite{Poke2016AllConcurLC}, \\ 
 & Multi-Paxos~\cite{10.1145/2673577} & Derecho~\cite{derecho}                         \\ \hline
 
    \multirow{2}{*}{\bf{Per-key order}}  & {\bf{CR}}~\cite{chain-replication}, CRAQ~\cite{10.5555/1855807.1855818},& {\bf{ABD}}~\cite{lynch:1997}, CP~\cite{10.1145/279227.279229},  \\ 
      & PB~\cite{primary-backup}, CHT~\cite{cht}  &  Hermes~\cite{Hermes:2020} \\
\hline
\end{tabular}
\end{center}
\caption{CFT protocols taxonomy. Using \projecttitle{}, we transform one protocol (shown in bold) of each category.} \label{tab:categories}
\vspace{-2pt}
\end{table}

\myparagraph{BFT protocols} In contrast to CFT protocols, BFT protocols assume very little about the nodes and the network; faulty nodes may behave arbitrarily while the network is unreliable. To tolerate $f$ arbitrarily faulty processes that may \emph{equivocate} (i.e., make conflicting statements for the same request to different replicas), BFT protocols add $f$ extra replicas to their system model requiring at least $3f+1$ replicas for safety. As such, BFT protocols exhibit worse scalability compared to CFT protocols (which only require at most $2f+1$ replicas).

BFT protocols are also limited in performance. They incur high message complexity ($O(f^2)$)~\cite{10.1145/2168836.2168866, Castro:2002, minBFT}, multiple protocol rounds~\cite{DBLP:journals/corr/LiuLKA16a, 10.1145/2168836.2168866, Castro:2002, DBLP:journals/corr/abs-1803-05069, yandamuri} and complex recovery ($O(f^2)$ in view-change)~\cite{DBLP:journals/corr/LiuLKA16a, Castro:2002, minBFT, 10.1145/2168836.2168866}. As an example of this, PBFT~\cite{Castro:2002}, a well-known BFT protocol, requires at least $3f+1$ nodes, executes three broadcast rounds, and incurs $O(n^2)$ message complexity. 


Thirdly, BFT protocols are complex, introducing burdens to developers. Guerraoui et al.~\cite{10.1145/2658994} found that most protocol implementations consist of thousands of lines of (non-trivial) code, e.g., PBFT~\cite{Castro:2002} and Zyzzyva~\cite{10.1145/1658357.1658358}. Even if system designers wish to use a state-of-the-art BFT protocol, optimizing it for the specific application settings (e.g., network bandwidth,  number of clients and replicas, cryptographic libraries, etc.) is a rather complicated task.  Even trivial changes or intuitive optimizations can be extremely hard and might affect other parts of the protocol (e.g., view-change in Zyzzyva).

\subsection{Transformation Requirements} The basic requirements for transforming a CFT protocol for Byzantine environments are established in a theoretical result published by Clement et al. in PODC 2012~\cite{clement2012}. This seminal paper shows that non-equivocation and transferable authentication are necessary to go from $3f+1$ to $2f+1$ replicas for a reliable broadcast in Byzantine settings. Our work shows that not only can this lower bound be achieved in practice, but we can do so while providing high performance by leveraging modern hardware in a cloud environment. Next, we discuss how \projecttitle{} satisfies these two fundamental requirements, while $\S$~\ref{sec:motivation} elaborates on how to design \em{practical} and \em{efficient} protocols that meet these requirements.

\noindent{}{\bf{\underline{Property 1:}}} The transferable authentication property refers to the authenticity of a received message, requiring that a replica must be able to verify that the supposed sender indeed had sent the message. The authentication is transferable if the original sender can be verified even for forwarded messages. Formally, a message $m$ received by a correct process $P_j$ from $P_i$ is verifiable by any other correct process $P_k$. That is, given an authentication proof $\sigma_i$: 
\[
\forall P_k : \text{Verify}(m, \sigma_i, P_i) \Rightarrow \text{Accept}(m, P_k)
\]

\noindent{}{\bf{\underline{Property 2:}}} The non-equivocation property guarantees that replicas cannot \emph{accept} conflicting statements for the same request. That implies that \projecttitle{} must detect attacks where adversaries try to compromise the protocol by sending invalid requests or by re-sending valid but stale requests (\emph{replay attacks}). Formally, a Byzantine node $P_i$ cannot produce two different messages (conflicting statements) $m$ and $m'$ for the same operation to different correct replicas $P_j$ and $P_k$: \[
\forall P_j, P_k, \quad (P_i \xrightarrow{} P_j : m) \wedge (P_i \xrightarrow{} P_k : m') \Rightarrow m = m'
\]

\subsection{Rethinking CFT Protocols} 
While Byzantine Fault Tolerant (BFT) protocols~\cite{Lamport:1982} offer important foundations for developing distributed systems with stronger guarantees in the presence of {\em Byzantine failures}, they are {\em not adopted \underline{in practice}} because of their high-performance and replication resource overheads, and implementation complexity~\cite{visigoth-eurosys}.

The “CFT vs. BFT" conundrum creates a fundamental design trade-off between the {\em efficiency of CFT protocols} for practical deployments and the {\em robustness of BFT protocols} for Byzantine settings of modern cloud environments. However, traditional BFT protocols design and evaluation has not taken into account {\em \underline{modern cloud hardware}}. 

\myparagraph{Key insights} Our work seeks to resolve this trade-off by leveraging modern cloud hardware to transform existing CFT protocols for Byzantine settings in untrusted cloud environments. Our transformation underpins the robustness and efficiency axes.
For {\em robustness}, we leverage trusted hardware available to harden the security properties of CFT protocols~\cite{intel-sgx, keystone_eurosys, amd-sev, intelTDX}. 

For {\em efficiency}, we leverage the modern networking hardware, such as RDMA/DPDK for kernel bypass, to design a highly optimized communication protocol for replicating the state across nodes in distributed settings~\cite{rdma, dpdk, erpc}, while overcoming the I/O bottlenecks in trusted computing~\cite{rkt-io}.


\myparagraph{Modern hardware in the context of BFT}
Trusted execution environments (TEEs)~\cite{cryptoeprint:2016:086, arm-realm, amd-sev, riscv-multizone, intelTDX} offer a hardware-enforced isolated computing environment that guarantees the integrity and confidentiality of its code and data, remaining resistant against all software attacks even in the presence of a privileged attacker (hypervisor or OS). 

In our work, we leverage trusted execution environments (TEEs) in the context of BFT by realizing the potential of TEEs in hardening the properties of CFT replication protocols in the presence of Byzantine actors (e.g., network adversaries, compromised OS/hypervisor, corrupted host memory, etc.) in the untrusted cloud.

High-performance distributed systems~\cite{fasst, farm} abandon the traditional kernel-based networking (sockets) to avoid syscalls' overheads~\cite{flexsc}. Instead, they adopt direct network I/O (RDMA~\cite{rdma}, DPDK~\cite{dpdk}) to map the device's address space into userspace,  bypassing the kernel stack.

We also adopt direct network I/O as it is even more well-suited to TEEs where syscall execution is extremely expensive~\cite{treaty, avocado}. We leverage eRPC~\cite{erpc}, a general-purpose and asynchronous remote procedure call (RPC) library for high-speed networking for lossy Ethernet or lossless fabrics. 

To sum up, we leverage TEEs and high-performant network stacks in the context of BFT to provide two key properties for successfully transforming a CFT protocol to operate in Byzantine settings, as identified by Clement et al.~\cite{clement2012}: (a) transferable authentication, i.e., the ability to establish trust in nodes in distributed settings by designing a remote attestation protocol, and (b) non-equivocation, i.e., once the trust is established in a node via the remote attestation protocol, the node follows the CFT replication protocol faithfully, and therefore, it cannot send conflicting statements to other nodes.

To realize the transformation requirements, we use modern hardware to implement the following mechanisms:

\noindent{}{\bf{Mechanism 1:}} We employ cryptographic primitives and an attestation protocol. The cryptographic primitives ensure that nodes can generate and validate authenticated messages while our attestation protocol ($\S$~\ref{sec:attestation}) ensures that only trusted replicas access the cryptographic keys and execute the protocol.

\noindent{}{\bf{Mechanism 2:}} We prevent equivocation by materializing a distributed TCB that shields the protocol's (distributed) execution as well as shielding the network communication based on an authenticated message format ($\S$~\ref{sec:normal_operation}).

\subsection{\projecttitle{} Overview}

\myparagraph{\projecttitle{} protocol overview} We consider a message-passing system consisting of $N$ nodes running the Recipe protocol. Among the $N$ nodes, $f$ exhibit Byzantine faults, while the remaining $N-f$ nodes are well-behaved. We assume $N \geq 2f + 1$. Well-behaved nodes follow the protocol faithfully, while malicious nodes may arbitrarily deviate. We define any group of $N-f$ or more nodes as a quorum.

\projecttitle{} executes a sequence of epochs, each potentially assigned a unique leader. Each epoch consists of several phases: Transferable authentication, Initialization, Normal operation (including request processing and commitment), View change, and Recovery.

The protocol is segmented into five key components:

\begin{enumerate}
    \item \textbf{Transferable authentication:} Before joining, each node undergoes a remote attestation procedure. The Protocol Designer (PD) interacts with a Configuration and Attestation Service (CAS) to verify the integrity of the node's Trusted Execution Environment (TEE). Only attested nodes receive configuration information and cryptographic keys.
    \item \textbf{Initialization:} Attested nodes establish secure communication channels and initialize their local key-value stores. The implemented CFT protocol (e.g., Raft) relies on its existing mechanisms to elect a leader before executing clients' requests.
    \item \textbf{Normal operation:}
    \begin{itemize}
        \item \textbf{Request processing:} Clients send attested requests $[\hc \sigmac, (metadata, req\_data)]$ to the leader. The leader verifies the request, "shields" it using cryptographic primitives within its TEE, creating $(\alphaa, \kv)$, and broadcasts it to the followers as $[\hr \sigmacq, (metadata', req\_data)]$, where $metadata'$ includes the view identifier $\view$, the communication channel identifier $\cq$ and the message sequence number identifier $\cntcq$ to prevent replay attacks. Followers execute the request within their TEEs and acknowledge the leader. 
        \item \textbf{Commitment:} A multi-phase commit protocol (dependent on the underlying CFT protocol) ensures that updates are consistently applied across all correct replicas.
    \end{itemize}
    \item \textbf{View change:} If the leader fails (detected via a trusted lease mechanism~\cite{t-lease} which is implemented as part of our secure runtime environment), a new leader is elected using the CFT protocol. Committed updates are preserved across view changes.
    \item \textbf{Recovery:} New or recovering nodes undergo the transferable authentication process and join the membership as {\emph fresh} replicas. They then synchronize their state with the existing replicas before fully participating.
\end{enumerate}

In a nutshell, \projecttitle{} guarantees safety by leveraging TEEs and cryptographic primitives to protect data and ensure that only valid operations are executed. The underlying CFT protocol and the trusted lease mechanism for failure detection provide liveness. The transferable authentication phase ensures that only authorized nodes participate in the protocol, preventing Sybil attacks~\cite{sybilAttack}.

\myparagraph{Formal analysis and verification} We formally verify the safety and security properties of \projecttitle{} using the symbolic model checker Tamarin~\cite{tamarin-prover}, assuming a Dolev-Yao attacker and perfect cryptography.
For this, we model an abstract \projecttitle{} setup as a labeled transition system. We can then verify a set of temporal properties on the transition traces of this system using Tamarin's automated deduction and equational reasoning. 
This allows us to verify the safety, integrity, and freshness properties that are presented in detail in \autoref{sec:formal-verif}.


\myparagraph{\projecttitle{} library} We materialise \projecttitle{} approach as a generic library, \projecttitle{}-lib ($\S$~\ref{sec:recipe-implementation}).
The \projecttitle{} library leverages TEEs along with direct I/O to resolve the tension between security and performance by building an efficient and practical transformation of unmodified CFT replication protocols for Byzantine settings. \projecttitle{} achieve this by implementing a distributed trusted computing base (TCB) that shields the replication protocol execution and {\em extends} the security properties offered by a single TEE (whose security properties are only effective in a single-node setup) to a distributed setting of TEEs. Our design is comprised of a transferable authentication phase ($\S$~\ref{sec:attestation}) for distributed trust establishment, a high-performant network stack for secure communication over the untrusted network ($\S$~\ref{subsec:networkin}) and a memory-efficient KV store ($\S$~\ref{subsec:KV}).


\myparagraph{\projecttitle{} evaluation} Our evaluation assesses \projecttitle{}'s generality and efficiency. Specifically, to show the generality of our approach, we apply and evaluate \projecttitle{} on real hardware with four well-known CFT protocols (from now on, an `R-' prefix stands for the transformed protocol); a decentralized (leaderless) linearizable multi-writer multi-reader protocol (ABD)~\cite{lynch:1997} (R-ABD), two leader-based protocols with linearizable reads, Raft~\cite{raft} (R-Raft) and Chain Replication (CR)~\cite{chain-replication} (R-CR), and AllConcur~\cite{Poke2016AllConcurLC} (R-AllConcur), a decentralized consensus protocol with consistent local reads. To evaluate performance, we compare \projecttitle{} protocols with two competitive BFT replication protocols, BFT-smart~\cite{bft-smart} (PBFT~\cite{Castro2002}), whose specific implementation has been adopted in industry~\cite{bftsmarthyperledger} and Damysus~\cite{10.1145/3492321.3519568} a state-of-the-art BFT replication protocol.  Our evaluation shows that \projecttitle{} achieves up to $24\times$ and $5.9\times$ better throughput w.r.t. PBFT and Damysus, respectively, while improving scalability---\projecttitle{} requires $2f+1$ replicas, $f$ fewer replicas compared to PBFT ($3f+1$). We further show that \projecttitle{} can offer confidentiality---a security property not provided by traditional BFT protocols---while achieving a speedup of $7\times$---$13\times$ w.r.t. PBFT and up to $4.9\times$ w.r.t. Damysus.

\subsection{Our Contributions} 
To summarize, we make the following contributions:

\begin{itemize}[leftmargin=*]
    \item {\bf Hardware-assisted transformation of CFT protocols:} We present \projecttitle{}, a generic approach for transforming CFT protocols to tolerate Byzantine failures without any modifications to the core of the protocols.

    \item {\bf Formal analysis and verification:} We formally verify the safety and security properties of \projecttitle{} using the Tamarin symbolic model checker~\cite{tamarin-prover}.  By modeling \projecttitle{}  as a labeled transition system and assuming a Dolev-Yao attacker~\cite{dolev-yao-model} with perfect cryptography, we verify key properties like safety, integrity, and freshness through automated deduction and equational reasoning. Therefore, we provide a correctness analysis for the safety and liveness properties of our
transformation of CFT protocols operating in Byzantine settings. 

    \item {\bf Generic \projecttitle{} APIs:} We propose generic \projecttitle{} APIs to transform the existing codebase of CFT protocols for Byzantine settings. Using \projecttitle{} APIs, we have successfully
transformed a range of leader-/leaderless-based CFT protocols enforcing different ordering semantics. 

    \item {\bf \projecttitle{} in action:} We present an extensive evaluation of \projecttitle{} by applying it to four CFT protocols: Chain Replication, Raft, ABD, and AllConcur. We evaluate these four protocols against the state-of-the-art BFT protocol implementations and show that \projecttitle{} achieves up to $24\times$ and $5.9\times$ better throughput.
    
\end{itemize}

\subsection{Software Artifact: Theory Meets Practice}
We have implemented the \projecttitle{} protocol as a generic library, which can be used to transform existing CFT protocols to operate in Byzantine cloud environments.   
\projecttitle{} will be available as an open-source project along with the entire experimental evaluation setup~\cite{receipe_artifact}. Our artifact includes the following:
 \begin{itemize}
 \item The \projecttitle{} library, including the secure runtime based on Intel SGX as our base TEE~\cite{intel-sgx}.

  \item The \projecttitle{} distributed data store architecture with the \projecttitle{}  replicated key-value store.

 \item Based on \projecttitle{}, a generic transformation of four CFT protocols: ABD~\cite{lynch:1997}, Raft~\cite{raft}, Chain Replication (CR)~\cite{chain-replication}, and AllConcur~\cite{Poke2016AllConcurLC}.

\item The formal verification proofs in Tamarin symbolic model checker~\cite{tamarin-prover}.

 \end{itemize}
 





\section{Related Work}
\label{sec:background}

Recall our research question:

\begin{tcolorbox}[colback=white, colframe=white, boxrule=0.1mm]
\emph{Can we leverage (and how) the advances in trusted computing (TEEs) and networking (direct I/O) to harden the properties of CFT protocols to (1) target a weaker fault model (i.e., Byzantine faults) while (2) offering performance and scalability? }
\end{tcolorbox}


Table~\ref{tab:recipe_vs_bft} compares the (most) related work with \projecttitle{} under those two axes: (1) their fault model, including the size of the implementation trusted computing base (TCB) and (2) performance or resource scalability including the number of active replicas, the message complexity as well as the use of novel hardware technologies (TEEs and D-IO). Resilience is the number of faulty nodes a protocol withstands (for safety and liveness). In contrast, direct-IO shows whether the protocols are implemented to work with direct network I/O, such as RDMA~\cite{rdma} or DPDK~\cite{dpdk}. 

We divide the related work that target BFT into two broad categories based on their resource scalability; the first includes classical BFT protocols that require (\emph{at least}) $3f+1$ participating replicas~\cite{DBLP:journals/corr/abs-1803-05069, bft-smart, Suri_Payer_2021, 10.1145/1658357.1658358, 6681599, all_about_eve} and the second category~\cite{10.1145/3552326.3587455, minBFT, hybster, DBLP:journals/corr/LiuLKA16a, A2M, 10.1145/2168836.2168866} refers to \emph{hybrid} BFT protocols that use trusted modules downgrading the replication degree to $2f+1$. In contrast to our \projecttitle{}, these approaches still require a working understanding of BFT; a task as challenging as it is error-prone~\cite{unsafeZyzzyva}.

\begin{table}[t]
\fontsize{7}{10}\selectfont 

\begin{center}
\begin{tabular}{ |c|c|c|c|c|c|c| } 
 \hline
 Protocols & Active/All Replicas & Resilience & Message complexity & TEEs/D-IO & Fault Model & TCB \\ [0.5ex] \hline \hline
 FastBFT~\cite{DBLP:journals/corr/LiuLKA16a}, CheapBFT~\cite{10.1145/2168836.2168866} & $f+1$/$2f+1$ & $0$ & $O(n), O(n^2)$ & Yes/No & Byz. & Small\\
 MinBFT~\cite{minBFT}, Hybster~\cite{hybster} & $2f+1$/$2f+1$ & $f$ &  $O(n^2)$ & Yes/No & Byz. & Small\\
 PBFT~\cite{Castro:2002}, HotStuff~\cite{DBLP:journals/corr/abs-1803-05069} & $3f+1$/$3f+1$ & $f$ & $O(n^2)$ $O(n)$ & No/No & Byz. & N/A\\
 CFT & $2f+1$/$2f+1$ & $f$ & depends on the protocol & No/Yes & Crash-stop. & N/A\\
 {\bf \projecttitle{}} & $\mathbf{2f+1}$/$\mathbf{2f+1}$ & ${\mathbf{f}}$ & depends on the protocol & \bf{Yes}/\bf{Yes} & \bf{Byz.} & \bf{Low}\\ 
 \hline
\end{tabular}
\end{center}
\caption{Replication protocols related work vs \projecttitle{}.}
\label{tab:recipe_vs_bft}
\vspace{0pt}
\end{table}




\myparagraph{Classical BFT protocols} In the first category, PBFT~\cite{Castro:2002} and its variations~\cite{CFF} run a three-phase protocol. Replicas broadcast messages and transit to the next phases after receiving {\emph{quorum certificates}}~\cite{Castro2002} from at least $2f+1$ distinct replicas leading to $O(n^2)$ message complexity. Zyzzyva~\cite{10.1145/1658357.1658358} offloads to the clients the responsibility to correct replicas' state in case of a Byzantine primary. However, prior work~\cite{unsafeZyzzyva} found safety concerns in the protocol.

Streamlined protocols~\cite{DBLP:journals/corr/abs-1803-05069, Chan2018PaLaAS, DBLP:journals/corr/abs-1807-04938, Chan2018PiLiAE} avoid heavy state transfers at the view-change by rotating the leader on each command at the cost of additional rounds. HotStuff~\cite{DBLP:journals/corr/abs-1803-05069} adds two extra phases to commit the latest blocks. Basil~\cite{Suri_Payer_2021} targets \emph{operability} when Byzantine nodes sabotage the execution requiring $5f+1$ replicas.

\myparagraph{Trusted hardware for hybrid BFT protocols} The second category includes {\emph{hybrid}} protocols~\cite{10.1145/3492321.3519568, minBFT, 10.1145/3552326.3587455, 10.1145/3492321.3519568, treaty, avocado, ccf} that leverage trusted hardware to optimize the performance of classical BFT at the cost of generalization and easy adoption. For example, MinBFT~\cite{minBFT} (a PBFT derivative optimized with TEEs), Damysus~\cite{10.1145/3492321.3519568} (a HotStuff derivative optimized with TEEs) and Hybster~\cite{hybster} use TEEs to decrease replication factor whereas others~\cite{levin2009trinc, 10.1145/1323293.1294280, yandamuri} utilize trusted counters and logs. Similarly, CheapBFT~\cite{10.1145/2168836.2168866} and FastBFT~\cite{DBLP:journals/corr/LiuLKA16a} build on trusted modules to use $f$+1 active replicas but transit to fallback BFT protocols in case of Byzantine failures. HotStuff-TPM~\cite{DBLP:journals/corr/abs-1803-05069} uses TPM~\cite{10.1109/SP.2010.32} at the cost of an extra phase. 

Similarly to \projecttitle{}, CCF~\cite{ccf} builds within a distributing setting of TEEs a variation of Raft consensus protocol (which operates under the CFT model). In contrast to \projecttitle{}, CCF builds a ledger, an append-only log, whereas \projecttitle{} is designed as a generic library to strengthen any CFT replication protocol that does not necessarily offer consensus.

\myparagraph{Programmable hardware for hybrid BFT protocols} Other works leverage programmable hardware, e.g. FPGAs~\cite{10.1145/2168836.2168866}, SmartCards~\cite{levin2009trinc} and switches~\cite{10.1145/3603269.3604874} to provide foundational primitives for ensuring BFT. For example, NeoBFT~\cite{10.1145/3603269.3604874} targets the BFT model for permissioned (BFT) blockchain systems~\cite{10.1145/3190508.3190538} by designing an \emph{authenticated ordered multicast} primitive in the programmable switch. To overcome the computation and scalability bottlenecks, they connect an FPGA device that serves as a cryptographic coprocessor to the switch. Compared to NeoBFT, where the switch is a single point of failure, \projecttitle{} offloads security into a distributed setting of TEEs, providing better availability guarantees. At the same time, it allows system designers to transform unmodified CFT protocols.  

Lastly, Trinc~\cite{levin2009trinc} and CheapBFT~\cite{10.1145/2168836.2168866} rely on peripherals to generate attestations for the exchanged message with extremely expensive latencies (50us---105ms)~\cite{levin2009trinc, 10.1145/2168836.2168866}. In addition, similarly to all previous hybrid protocols, they are specifically designed to optimize a particular variation of a BFT system and do not offer a generic methodology for any replication protocol.

\section{\projecttitle{} Protocol}
\label{sec:recipe-protocol}

We first describe the system model. Next, we define the primitives of \emph{non-equivocation and authentication} as well as \emph{attestation} that we use to build our \projecttitle{} protocol. Based on these foundations, we present the \projecttitle{} protocol.

\subsection{System Model}
\label{sec:system_model}
\myparagraph{Replication protocol}
A replication protocol ensures that a set of $N$  replicas $(R_i,...,R_n)$ maintains a consistent and available state $S$ despite failures or concurrency. A replication protocol operates over a distributed state machine, where replicas execute a deterministic sequence of operations from the set of operations $O$, s.t. $S \times O \rightarrow S$, and apply them in a consistent order.
Since not all replication protocols solve consensus and thus are not state-machine-replication (SMR) protocols, we do not assume the typical consensus properties (termination, agreement, and validity). Instead, we assume that the replication protocols should guarantee the following properties:

\begin{itemize}
    \item \textbf{Consistency:} All correct replicas agree on the same sequence of operations.
   \item \textbf{Availability:} Replicas process requests as long as $<f$ replicas fail.
    \item \textbf{Fault tolerance:} The system operates correctly with up to $f$ failures (crash or Byzantine).    
\end{itemize}

\myparagraph{Model sketch}
We model the distributed system as a set of $N$ TEEs in $N$ nodes (or replicas), each hosting either a \emph{follower} or a \textit{coordinator} process $P_i$ which executes a CFT protocol. The system is modeled as a state machine, where each replica $R_i$ maintains a local state $S_i$ and transitions between states based on received messages and protocol execution rules.
We assume that \projecttitle{}'s nodes run in a third-party untrusted cloud infrastructure. 
A coordinator serves client requests by initiating the implemented CFT replication protocol. Upon completion, it replies back to clients. In leaderless protocols, coordinators are selected randomly (any node can act as a follower and/or a coordinator). In leader-based protocols, only the active leader can act as a coordinator, the rest of the nodes are followers. 

\myparagraph{Communication model} 
Nodes communicate via a fully-con\-nected, bidirectional, point-to-point and unreliable message-passing network, where messages can be arbitrarily delayed, re-ordered or dropped. In line with previous BFT protocols, we adopt the partial synchrony model~\cite{10.1145/42282.42283}, where there is a known bound $\Delta$ and an unknown Global Stabilization Time (GST), such that after GST, all communications arrive within time $\Delta$.


\myparagraph{Fault and threat model}
We say that a node $N_i$ is \textit{faulty} if it exhibits Byzantine behavior~\cite{Lamport:1982}. 
The unprotected (\emph{out-of-the-TEE}) infrastructure (e.g., host memory, OS, NIC, network infrastructure/adversaries) can exhibit arbitrary Byzantine behavior while we assume that the TEEs can only crash-fail. We say that a node is faulty if one of the following holds true:
\textit{(i)} the TEE fails by crashing or \textit{(ii)} the unprotected infrastructure is Byzantine.
Safety is defined as follows:  If a correct replica $R_j$ delivers a message $m$  from $R_i$, then $R_i$  must have previously sent $m$, and $m$ is consistent with the protocol state. Liveness is defined as follows: If a client submits a request $r$, and a majority of replicas are correct, then $r$ is eventually committed. Last, for safety and liveness, we assume that for $N \ge (2f+1)$ nodes up to $f$ can be \emph{faulty}.

\myparagraph{Cryptographic model} 
We assume collision resistance for the hash functions; no computationally bounded adversary can find two distinct inputs $m \neq m'$ such that $hash(m)=hash(m')$, except with negligible probability. We also make the conventional assumption that signatures and keys are unforgeable, the initial keys are generated securely, and the private keys are stored securely in the TEE.

\subsection{\projecttitle{} Primitives}
\myparagraph{Non-equivocation and authentication primitives} \projecttitle{}'s embodies a non-equivocation and an authentication layer through two TEE-assisted primitives, the shield\_request() and verify\_request(), shown in Algorithm~\ref{algo:primitives}.

\begin{algorithm}[t]
\SetAlgoLined
\small


$\triangleright$ cnt$_{cq}$: the latest sent message id from  cq\\$\triangleright$ rcnt$_{cq}$: the last committed message id from cq


\textbf{function} shield\_request(req, cq) \{ \\
\Indp
cnt$_{cq}$ $\leftarrow$ cnt$_{cq}$+1; t$\leftarrow$ (view, cq, cnt$_{cq}$);\\
$[$$h_{\sigma_{cq}}$, (req,t)$]$  $\leftarrow$ singed\_hash(req, t);\\
\textbf{return} $[$$h_{\sigma_{cq}}$, (req,t)$]$;\\
\Indm
\} \\

\textbf{function} verify\_request($h_{\sigma_{cq}}$, req, (view, cq, cnt$_{cq}$)) \{ \\
\Indp
    \textbf{if} verify\_signature($h_{\sigma_{cq}}$, req, (view, cq, cnt$_{cq}$)) == True \textbf{then}\\
    \Indp
        \textbf{if} view == current\_view \textbf{then}\\
        \Indp
            \textbf{if} cnt$_{cq}$ <= rcnt$_{cq}$ \textbf{then}\\
            \Indp
                \textbf{return} [False, req, (view, cq, cnt$_{cq}$)]; \\
            \Indm
            \textbf{if} cnt$_{cq}$ == rcnt$_{cq}$+1 \textbf{then} rcnt$_{cq}$ $\leftarrow$ rcnt$_{cq}$+1;
            buffer\_locally(req, (view, cq, cnt$_{cq}$));\\
                \textbf{return} [True, req, (view, cq, cnt$_{cq}$)]; \\
            
        \Indm
    \Indm
    \textbf{return} [False, req, (view, cq, cnt$_{cq}$)]; \\

\Indm
\} \\
\vspace{-1pt}
\caption{\projecttitle{}'s authentication primitives.}
\vspace{-3pt}
\label{algo:primitives}
\end{algorithm}

\noindent{Non-equivocation layer}: \projecttitle{} prevents replay attacks in the network with sequence numbers for the exchanged messages. Each replica maintains local sequence tuples $t = (view, cq, cnt_{cq})$ where $view \in \mathbb{N}$ is the current view number, $cq$ is the communication endpoint(s) between two nodes, and $cnt_{cq}$ is the current trusted counter value in that view for the latest request sent over $cq$. The sender assigns a unique tuple to messages and then increments the trusted counter to guarantee monotonicity. This guarantees monotonicity: $ \forall m_j^y, m_j^x$, if $y>x$, then $cnt_{cq}(m_j^y) > cnt_{cq}(m_j^x)$.
Replicas execute the implemented CFT protocol for verified valid requests. Replicas can verify the freshness of a message by examining its cnt$_{cq}$ (verify\_request() primitive). The primitive verifies that the message's id (as part of the metadata) is consistent with the receiver's local counter rcnt$_{cq}$ (rcnt$_{cq}$ is the last seen valid message counter for received messages in cq). \projecttitle{}'s replicas are willing to accept ``future'' valid messages as these might come out of order, i.e., $cnt_{cq} > (rcnt_{cq}+1)$. These messages are processed and committed according to the CFT protocol.

\myparagraph{Authentication primitive} For the authentication, we use cryptographic primitives (e.g., MAC and encryption functions when \projecttitle{} aims for confidentiality) to verify the integrity and the authenticity of the messages. Each message $m$ sent from a node $n_i$ to a node $n_j$ over a communication channel $cq$ is accompanied by metadata (e.g., $cnt_{cq}$, view, sender and receiver nodes id), and the calculated message authentication code (MAC) $h_{cq_{\sigma_q}}$. The MAC is calculated over the payload and the metadata, then follows the message $m$. Formally, $m \leftarrow \langle req, (view,cq,cnt_{cq}),h_{cq_{\sigma_q}}(req ||view || cq || cnt_{cq})) \rangle$. The sender node calls into the shield\_request(req, cq) and generates such a trusted message for the request req. On the receiver side, $Accept(ID_j,\langle req,(view,cq,cnt_{cq}),\sigma\rangle) \Longleftrightarrow Verify(h_{cq_{\sigma_q}}, req, (view || cq ||cnt_{cq})).$

\myparagraph{Attestation primitive} Remote attestation is the building block to verify the authenticity of a TEE, i.e., the code and the TEE state are the expected~\cite{Parno2010}. As such, \projecttitle{} provides attest(), generate\_quote() and remote\_attestation() primitives (Algorithm~\ref{algo:attestation}) that allow replicas to prove their trustworthiness to other replicas or clients. The attestation takes place before the control passes to the protocol's code. Only successfully attested nodes get access to secrets (e.g., signing or encryption keys, etc.) and configurations. 


\begin{algorithm}[t]
\SetAlgoLined
\small

\textbf{function} remote\_attestation() \{ \\
 \Indp
 nonce $\leftarrow$ generate\_nonce();\\
 \textbf{send}(nonce, k$_{pub}$); \textbf{DHKE}(); quote$_{\sigma_{k_{pub}}}$ $\leftarrow$ \textbf{recv}();\\
 \textbf{if} verify\_signature(quote$_{\sigma_{k_{pub}}}$) == True \textbf{then}\\
    \Indp
        $\mu_{TEE}$ $\leftarrow$ decrypt(quote$_{\sigma_{k_{pub}}}$, k$_{priv}$);\\
        \textbf{if} (verify\_quote$(\mu_{TEE})$ == True) send\_secrets();\\
    \Indm
 \Indm
\} \\

\textbf{function} attest() \{ \\
\Indp
    $\mu$ $\leftarrow$ gen\_enclave\_report(); \textbf{return} $\mu$;\\ 
\Indm
\} \\

\textbf{function} generate\_quote($\mu$, k$_{pub}$) \{ \\
\Indp
    key$_{hw}$ $\leftarrow$ EGETKEY();\\
    quote $\leftarrow$ sign($\mu$, key$_{hw}$); 
    quote$_{\sigma_{k_{pub}}}$ $\leftarrow$ sign(quote, k$_{pub}$);\\
    \textbf{return } quote$_{\sigma_{k_{pub}}}$;\\
\Indm
\} \\
\caption{\projecttitle{}'s attestation primitive.}
\label{algo:attestation}
\vspace{-3pt}
\end{algorithm}


\subsection{Clients in \projecttitle{}}
Clients in \projecttitle{} execute requests through a \texttt{PUT}/\texttt{GET} API. As discussed in $\S$~\ref{subsec:overview}, the request is forwarded to the protocol's coordinator node. Figure~\ref{fig:raft_example} shows as an example a \projecttitle{} implementation of Raft (R-Raft) including all three execution phases of a typical \projecttitle{} protocol: the transferable authentication phase (blue box), the initialization phase (green box) and the normal execution phase where the transformed CFT protocol executes clients' requests (red box). Prior to the protocol execution, nodes pass through a transferable authentication phase ($\S$~\ref{sec:attestation}) to prove that the TEEs and loaded code are genuine, followed by initialization and normal operation.

\subsection{Normal Operation}
\label{sec:normal_operation}

We first explain the initialization and the normal execution phases, assuming all participant nodes executed the transferable authentication phase successfully.
The nodes execute the initialization phase, initializing their own local KVs \circled{B.1} and their network connections (e.g., configures NIC-memory, network ports, etc.) and establish connections with other peers \circled{B.2} based on the configuration it securely received at the attestation process \circled{A.7}.

The leader then runs the underlying CFT protocol (in our case, Raft \circled{C.1}---\circled{C.9}) to execute the client request \circled{R.1}. Upon completion, it replies back to the client \circled{R.2}. Next, we discuss the \projecttitle{} abstraction under the normal operation.

\begin{figure*}
    \begin{center}
        \includegraphics[width=\textwidth]{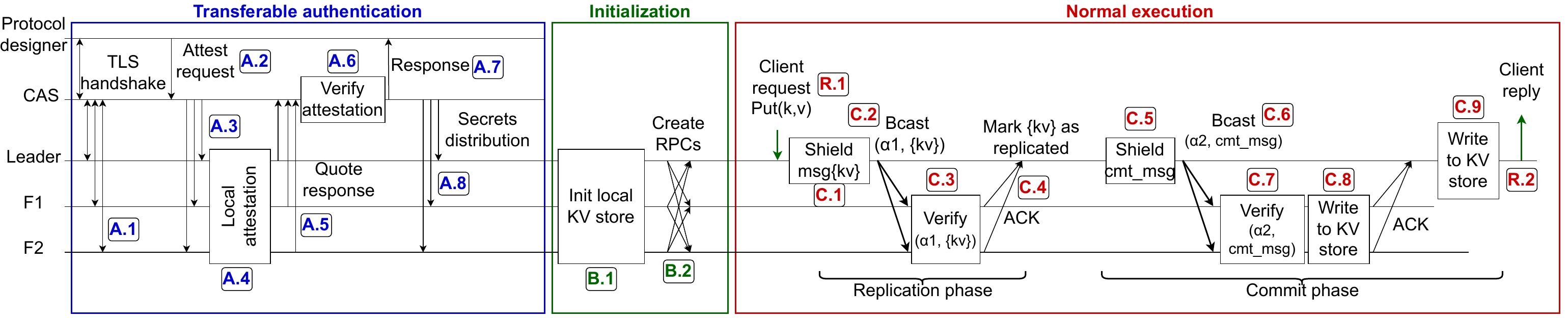}
        \vspace{-2pt}
        \caption{Example of the \projecttitle{} version of Raft (R-Raft) execution.}
        \label{fig:raft_example}
    \end{center}
\end{figure*}

We use the notation [$h_{c_{\sigma_c}}$, payload] to denote an \emph{attested} or \emph{shielded}  message that is comprised of the signed hash ($h_{c_{\sigma_c}}$) of payload (\emph{certificate}) along with the raw payload data. We use the symbol $\sigma_{c}$ to denote that a piece of data is signed with a key $c$. Figure~\ref{fig:raft_example} uses the notation ($\alpha$, {kv}) for an attested message referring to a key-value pair kv with certificate $\alpha$.



    \noindent{}{\bf{\underline{\#1}}:} Clients send the coordinator their request of the form [$h_{c_{\sigma_c}}$, (metadata, req\_data)] \circled{R.1}. The req\_data is the request's associated data and the metadata might include among others the client's and the request's id, the leader's and term's ids (known to the client). 
    
    
    \noindent{}{\bf{\underline{\#2}}:} Nodes receive and process a request after successfully verifying their integrity and authenticity. \projecttitle{}'s protocols inherit the constraints of the original CFT protocol. For instance, our R-Raft leader will drop requests with the wrong view of the term or leader.
    
    \noindent{}{\bf{\underline{\#3: Upon the reception of a client's request:}}}
    
    {\bf{\#3.1}} The coordinator (leader) verifies the integrity and authenticity of the message using \projecttitle{}'s authentication layer. It also verifies the metadata, e.g., the message is invalid if the term and the leader (if any) known to the client are incorrect. The leader updates the client table with the latest processed request for each client. 
    
    {\bf{\#3.2}} Next, the leader initializes the protocol for that request. In our example, the Raft leader shields the message \circled{C.1}, generating a trusted message format ($\alpha1$, {kv}) where $\alpha1$ is the certificate of {kv} and broadcasts the request to the followers \circled{C.2} (replication phase).  

    {\bf{\#3.3}} The messages exchanged between replicas are of the form [$h_{r_{\sigma_{cq}}}$, (metadata, req\_data)]. The metadata includes a per-request unique tuple (view, cq, cnt$_{cq}$) that contains: (1) the view, an identifier that is optionally set by the implementation for every new leader (if any) (2) the communication channel id (cq) and (3) a sequencer id or a message counter (cnt$_{cq}$) that is assigned to the messages sent over this channel and is increased monotonically for every new message.
    
    \noindent{}{\bf{\underline{\#4: When a replica receives a message:}}}

    {\bf{\#4.1}} If the replica is in normal state operation, it verifies the message's validity. Else, it refuses to process the request.
    
    {\bf{\#4.2}} The replica verifies the received sequencer id (recv$_{cnt}$) to see if it is consistent with its local counter (cnt$_{cq}$). If $recv_{cnt} = (cnt_{cq}+1)$, the replica executes the request immediately, increases its local counter, acknowledges the sender node, and updates the client table. If the recv$_{cnt}$ refers to a ``future'' message (recv$_{cnt}$$>$cnt$_{cq}$+$1$), the replica queues the request in the protected area. Periodically, it applies the queued requests eligible for execution and notify coordinators accordingly.

    \noindent{}{\bf{\underline{\#5}}:} In our example, the followers verify the request \circled{C.3}, enqueue the un-committed request in a TEE buffer, and send ACKs back to the leader. The leader, upon receiving the majority of ACKs marks the request as replicated \circled{C.4} and proceeds to the second round of the protocol instructing the followers that replied to commit the update (\circled{C.5}---\circled{C.7}). At this point, each follower instructed to commit applies the request to its local KV store \circled{C.8} and ACKs the commit to the leader. Similarly to the replication phase, the leader finally commits \circled{C.9} when it receives ACKs from the majority.
    
    \noindent{}{\bf{\underline{\#6}}:} After the protocol's execution, the coordinator marks the request as committed and notifies the client \circled{R.2}.
    

\subsection{View Change}

While decentralized protocols remain available as long as most nodes are part of the membership, the leader-based protocols do not progress if the leader goes down. To remain available after the leader crashes, the followers need to closely monitor the leader (e.g., heartbeat messages in inactive periods) and, in case it is unreachable, to \emph{elect} a new one, i.e., perform a \emph{view change}.



In line with the CFT protocols, \projecttitle{} protocols must assign a leader with a term and a node identifier. The term id can be seen as an \emph{epoch}, a monotonically increasing counter that uniquely identifies the current view of the system. To continue serving requests after a leader election, the majority must confirm the new leader and the new term. Since a leader needs to be acknowledged by the majority of the nodes to operate, the latest term will survive in at least one node, ensuring the term's monotonic increments.

\myparagraph{Correctness} The correctness condition for leader elections is that every commit must survive into the new leader election in the order selected for it when it was executed. \projecttitle{} does not make further assumptions, protocols can rely on their election algorithms as we guarantee the replicas are trusted.

\myparagraph{Failure detection} CFT protocols~\cite{raft, chain-replication} often require trusted timers to detect failures. \projecttitle{} builds on top of Intel SGX, which does not secure timers~\cite{monotoniccounterssgx, sgxtrustedtime}, whereas OS-timers and software clocks cannot be trusted. To overcome this, \projecttitle{} implements a trusted lease mechanism~\cite{t-lease}. Our mechanism supports all the properties of classical leases~\cite{10.1145/74850.74870}  that are the building block for trusted timeouts, failure detectors~\cite{222603}, leader election~\cite{815321}, etc.


\subsection{Transferable Authentication}
\label{sec:attestation}
Before initialization, all participant nodes run the transferable authentication phase (are \emph{attested}). The phase ensures that only authenticated replicas receive configurations and secrets and participate to the protocol, guaranteeing the transferable authentication property and protecting against Sybil attacks~\cite{10.1007/3-540-45748-8_24}. \projecttitle{} materializes this phase using a remote attestation protocol.

The attestation protocol is initialized by the protocol designer (PD) (\emph{challenger}), who establishes a TLS connection with the Configuration and Attestation service (CAS) \circled{A.1}. CAS is responsible for proving the authenticity of a TEE. For now, we focus solely on the attestation protocol; the CAS is discussed in $\S$~\ref{subsec:attestation}. The CAS also establishes secure communication channels with the participant nodes. 

The PD sends an \emph{attest request} to the CAS \circled{A.2}, which is then forwarded to the replicas \circled{A.3}. The replicas perform \emph{local attestation}~\cite{Parno2010}, i.e., they calculate a measurement of their code and generate a quote that is uniquely bound to that particular TEE \circled{A.4}. The quotes are sent over the TLS channel to the CAS for verification. Upon a successful attestation of a TEE, the CAS notifies the PD to forward configurations to the replicas \circled{A.7}---\circled{A.8}.

\subsection{Recovery}
\label{sec:recovery}
As nodes fail, new or recovered nodes need to be added to continue operating at peak performance. To add a new node, the membership needs to be reliably updated following the notification of all other live replicas of the new node’s
intention to join the replica group. For non-equivocation, recovered nodes always
start as fresh nodes and as such are assigned unique node ids by the CAS
through the attestation phase. Overall, a new joining node follows the next steps:


   \noindent{}{\bf{\underline{\#1}}:} A recovering node needs first to be attested before any secrets and membership information are shared. Before the control passes to the CFT protocol, the node sends a join request to a designated node, notifying it about its willingness to join the cluster. 
   
   \noindent{}{\bf{\underline{\#2}}:} The challenger-node that receives the request initializes a remote attestation to verify the new node's trustworthiness ($\S$~\ref{sec:attestation}). 
   
    \noindent{}{\bf{\underline{\#3}}:} After a successful attestation, as a response to the join-request, the challenger-node shares the network signing or encryption keys and the configuration of the membership. The challenger-node also broadcasts a message to the other replicas about the successful attestation of the new node. Once the new joiner acknowledges the response from the challenger-node, it establishes connections with the other replicas. 
    
    
    \noindent{}{\bf{\underline{\#4}}:} The new node joins as a shadow replica fetching the state of the system as in~\cite{10.1145/2815400.2815425, 10.1145/2043556.2043560}. If the CFT protocol allows, this node can participate in writes while recovering. Once synchronized with the system's state, it transitions to normal protocol operation. 


\section{\projecttitle{} Analysis and Formal Verification}
\label{sec:theory}

\subsection{Requirements Analysis}
We show how \projecttitle{} satisfies the non-equivocation and the transferable authentication properties.

\myparagraph{Non-equivocation}
\projecttitle{} prevents equivocation attacks through a trusted monotonically increasing message counter $cnt_{cq}$ that assigns \emph{sequence numbers} to the network messages. The sender assigns a monotonically increasing sequence number to every message of a given round, $cnt_{cq} \leftarrow cnt_{cq} +1$, guaranteeing a total ordering of all network messages between any two communication endpoints. Formally, for any two messages $m1,m2$ over a communication channel $cq$: $cnt_{cq}(m_{1}) < cnt_{cq}(m_2) \Longrightarrow  m_{1} \prec m_2$. 
On the receiving side, it suffices for replicas to verify that the message's counter is \emph{consistent} with their local known sequencer for this communication endpoint, $cnt_{cq} = rcnt_{cq} +1$. 
\projecttitle{}'s sequencer prevents the replays (stale but authenticated messages), which is indistinguishable from equivocation, $cnt_{cq} < rnt_{cq}$. In addition, a Byzantine node may ``appear'' to not send messages to some (weak non-equivocation) or all (strong non-equivocation) other nodes during a given operation~\cite{madsen2020}. \projecttitle{} is responsible for neither---we rely on the CFT as both weak and strong non-equivocation are indistinguishable from crash failures~\cite{madsen2020}.


\myparagraph{Transferable authentication} 
\label{transferable-authentication}
\projecttitle{} ensures the following two core properties from its TEE-assisted primitives: property \#1: \projecttitle{} distributes the configuration, keys etc. in a secure manner to trusted nodes, and property \#2: \projecttitle{} preserves the authenticity and integrity of the network messages.

Transferable authentication is provided implicitly by properties \#1 and \#2. 
 Property \#1 ensures that their signing keys are shared for every communicating pair of processes after their successful attestation. Recall that configuration data, signing keys, and other secrets are securely provisioned only to trusted nodes that have successfully completed remote attestation. This, in turn, follows that only trusted (correct) processes can sign (and generate) valid messages, since every message $m$ is signed by the sender's TEE using a private key $sk$. It also follows that Byzantine adversaries cannot alter or forge messages without the signing key, including ``future'' messages; instead, they are only limited to replaying old messages. Formally, $Pr[Verify(\sigma_c,m,k_{pub})=1] \leq negl(\lambda)$, where $\sigma$ is the signature, $\lambda$ is the security parameter (e.g., key size), and $negl(\lambda)$ is a negligible function, meaning it decreases faster than any polynomial function. Lastly, authenticity is transferable and can be verified in the exact same way that any two directly communicating nodes do.

\subsection{Correctness Analysis}
CFT protocols need to provide the following safety properties regarding the messages delivered by the network~\cite{making_distributed_app_rob, 268272}. We show how these are provided by \projecttitle{}'s non-equivocation and (transferable) authentication layers.


\noindent{\bf{Safety.}} If a correct process $p_i$ \emph{receives and accepts} a message $m$ from a process $p_j$, then the sender $p_j$ is correct and has executed the \texttt{send} operation with $m$.

\noindent{\bf{Integrity.}} If a correct process $p_i$ \emph{receives and accepts} a message $m$, then $m$ is a valid and correct message generated according to the protocol specifications.

\noindent{\bf{Freshness.}} If a correct process $p_i$ \emph{receives and accepts} a valid message $m_{j_x}$ sent from a correct process $p_j$, then it will not accept any future message $m_{j_y}$ with the same identifier, $y = x$, $\forall$$x,y \in \mathbb{N^+}$.


Next, we explain how \projecttitle{} satisfies these properties. 
Safety and integrity are directly satisfied by our transferable authentication mechanisms. Firstly, every message $m$ is signed by the sender’s TEE using a private key $k_{priv}$, and receivers verify $m$ using the sender’s public key $k_{pub}$. Thus, only trusted and correct processes can generate valid messages (i.e., valid signatures) that can be successfully verified: a message \textit{m} accepted by some correct process $p_i$ must have been generated and sent by a correct process $p_j$. Moreover, correct processes cannot deviate from the protocol's specification to generate messages that do not adhere to it.  Byzantine adversaries can neither forge nor alter messages without $k_{priv}$ ($\S$ \ref{transferable-authentication}).


Freshness is directly satisfied by our non-equivocation layer that by using monotonically increasing counters, imposes a total order on messages between two communication endpoints. A correct process $p_i$ drops already received messages to sustain replay equivocation attacks. 





\subsection{Formal Verification of the \projecttitle{} Protocol}~\label{sec:formal-verif}
We formally verify the previously mentioned safety and additional security properties of \projecttitle{} using Tamarin~\cite{tamarin-prover}. Tamarin operates in the symbolic (Dolev-Yao) model~\cite{dolev-yao-model} and thus requires us to make the following assumptions: (1) we do not consider individual bits, but instead atomic terms, like a counter, cryptographic key, etc., which are composed to derive messages (2) all cryptographic functions are pure, i.e., they have no side effects, and perfect (e.g., no hash collision) (3) a potential attacker can read and delete all messages sent on the network and modify them using the functions built into Tamarin or explicitly provided in the model (e.g., no side channels). Based on these assumptions, we can model the system state as a multiset, and the possible state transitions as multiset rewriting rules, resulting in a labeled transition system. In the case of \projecttitle{}, this involved mapping of the transferable authentication, initialization, and execution phase to \textit{facts} stored in the multiset and \textit{rules} for each operation that modifies the system state. This allows us to consider an unbounded number of processes, messages, and protocol executions. We utilize the results of previous TLS verification work~\cite{tls-proof1,tls-proof2,tamarin-tls-proof}, to abstract the TLS handshake and subsequent connection as a secure channel in our model.

In order to verify the properties of this system, we annotate rules with parameterizable \textit{action facts} and use them to express temporal first-order properties on all possible \textit{traces}, i.e., transition sequences. Tamarin can then use deduction and equational reasoning to derive either a proof of correctness or a counterexample, which violates our property~\cite{tamarin-prover}. To express the temporal relation of action facts we will use \(a\at{}t_i\), to express that action fact \(a\) occurred in the trace at time point \(t_i\). The relation \(t_a \prec t_b\) specifies that \(t_a\) occurred strictly before \(t_b\) in the trace, and \(t_a \equiv t_b\) expresses that both occur at the same time, which implies that they map to the same rule execution. Using this framework, we can express the following safety properties:

We use the action facts \textit{Acc}(ept) and \textit{Send}, which map to the according process operations, as well as the action fact \textit{Tr}(usted), which marks the rule where the processes are finally attested and from which point onward they are trusted. The first property we verify, which corresponds to our safety and integrity properties, is that sent and accepted (by some process) messages always originate from a trusted (correct) process or formally in Tamarin:

\begingroup
\setlength{\abovedisplayskip}{0.0em} 
\setlength{\belowdisplayskip}{0.4em} 
\setlength{\jot}{0.1em} 

\begin{equation}
    \begin{split}
    \forall~ p_i, m_{j_x}, t_{t_i}, t_a ~:&~ Tr(p_i) \at{} t_{t_i} \;\wedge\; Acc(p_i, m_{j_x}) \at{} t_a \;\wedge\; t_{t_i} \prec t_a \\
    \implies & \exists~ p_j,  t_{t_j}, t_s ~:~ Tr(p_j) \at{} t_t \;\wedge\; Send(p_j, m_{j_x}) \at{} t_s \;\wedge\; t_{t_j} \prec t_s \prec t_a
    \end{split}
\end{equation}
\endgroup

\noindent{We define property (2) to verify that messages are always accepted in the order they are sent. We express this in Tamarin as:}




\begingroup
\setlength{\abovedisplayskip}{0.0em} 
\setlength{\belowdisplayskip}{0.4em} 
\setlength{\jot}{0.1em} 

\begin{equation}
    \begin{split}
    \forall~ p_i, m_{j_x}, m_{j_y}, t_{t_i}, t_{a_x}, t_{a_y} ~:&~ Tr(p_i) \at{} t_{t_i} \;\wedge\; Acc(p_i, m_{j_x}) \at{} t_{a_x} \;\wedge\; Acc(p_i, m_{j_y}) \at{} t_{a_y} \;\wedge\; t_{t_i} \prec t_{a_x} \prec t_{a_y} \implies \\
    & \exists~ p_j, t_{s_x}, t_{s_y} ~:~ Send(p_j, m_{j_x}) \at{} t_{s_x} \;\wedge\; Send(p_j, m_{j_y}) \at{} t_{s_y} \;\wedge\; t_{s_x} \prec t_{s_y} \\
    \end{split}
\end{equation}
\endgroup

\noindent{Finally, we verify that messages are never accepted twice, which corresponds to our freshness property, expressed in Tamarin as:}

\begingroup
\setlength{\abovedisplayskip}{0.0em} 
\setlength{\belowdisplayskip}{0.4em} 
\setlength{\jot}{0.1em} 

\begin{equation}
    \begin{split}
    \forall~ p_i, m_{j_x}, m_{j_x}, t_{t_i}, t_{a_x}, t_{a_y} ~:&~ Tr(p_i) \at{} t_{t_i} \;\wedge\; Acc(p_i, m_{j_x}) \at{} t_{a_x} \;\wedge\; Acc(p_i, m_{j_x}) \at{} t_{a_y} \\
    &\;\wedge\; t_{t_i} \prec t_{a_x}, t_{a_y} \implies t_{a_x} \equiv t_{a_y}
    \end{split}
\end{equation}
\endgroup

\noindent{All of these properties are successfully verified by Tamarin for any number of TEEs, processes, and protocol executions, even in the presence of an attacker, as outlined above. A link to the full Tamarin model and proof artifact, which consists of more than seven thousand generated lines, will be provided after the double-blind review.}  

\section*{Appendix: \projecttitle Architecture, Implementation, and Evaluation}
\appendix
\section{\projecttitle Library}
\label{sec:abstraction}
\label{sec:recipe-implementation}

\subsection{\projecttitle{} Architecture Overview} \label{subsec:overview}
\myparagraph{Distributed systems architecture} 
A distributed data store system uses a tiered architecture with a distributed data store layer for routing requests, a replication layer (provided by \projecttitle{}) for consistent data replication, and a data layer (Key-Value stores) for storage.  \projecttitle{} provides a trusted computing base using trusted execution environments (TEEs) to protect consensus fault tolerance protocols, including secure initialization of replicas and a direct I/O layer for efficient, secure communication.

\begin{figure*}[t]
    \begin{center}
        \includegraphics[width=1\textwidth]{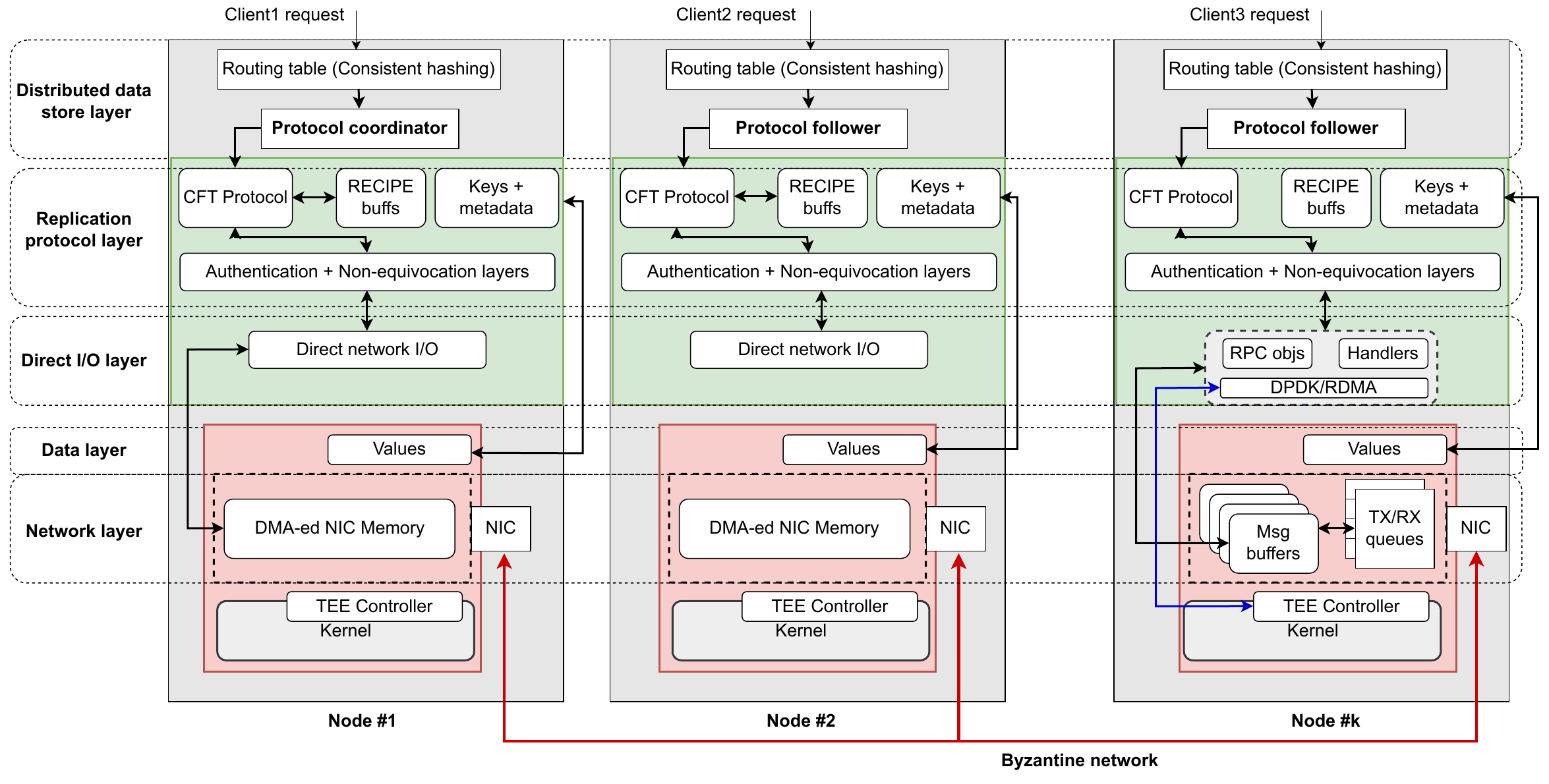}
        \caption{\projecttitle{}'s system architecture.}
        \label{fig:overview}
    \end{center}
\end{figure*}

Figure~\ref{fig:overview} shows the overview of a distributed data store system that builds on top of the \projecttitle{} system.  Distributed data stores implement a tiered architecture consisting of a \emph{distributed data store layer}, \emph{replication layer}, and  \emph{data layer}. In our case, the replication and data layers are provided by \projecttitle{}. The distributed data store layer maintains a routing table that matches the keyspace with the owners' nodes. This layer is responsible for forwarding client requests to the appropriate coordinator nodes (e.g., leader of the replication protocol) for execution. The \projecttitle{} replication layer is responsible for consistently replicating the data by executing the implemented protocol. After the protocol execution, \projecttitle{} nodes store the data in their Key-Value stores (KVs), the data layer, and they reply to the client~\cite{redis, rocksdb, leveldb, memcached2004}.

\myparagraph{\projecttitle{} architecture} \projecttitle{} design is based on a distributed setting of TEEs that implement a (distributed) trusted computing base (TCB) and shield the execution of unmodified CFT protocols against Byzantine failures. \projecttitle{}'s TCB contains the CFT protocol's code along with some metadata specific to the protocol. 

The code and TEEs of all replicas are attested before instantiating the protocol to ensure that the TEE hardware and the residing code are genuine. All authenticated replicas receive secrets (e.g., signing or encryption keys) and configuration data securely at initialization. 

Further, \projecttitle{} builds a \emph{direct I/O layer} comprised of a networking library for low-latency communication between nodes ($\S$~\ref{subsec:networkin}). The library bypasses the kernel stack for performance and shields the communication to guarantee non-equivocation and transferable authentication against Byzantine actors in the network. \projecttitle{} guarantees both properties by layering the non-equivocation and authentication layers on top of the direct I/O layer. In addition,  to strengthen \projecttitle{}'s security properties and eliminate syscalls, we map the network library software stack to the TEE's address space.

Lastly, \projecttitle{} builds the  \emph{data layer} on top of local KV store instances. Our design of the KV store increases the trust in individual nodes, allowing for local reads ($\S$~\ref{subsec:KV}). Our KV store achieves two goals: first, we guarantee trust to individual replicas to serve reads locally, and second, we limit the TCB size, optimizing the enclave memory usage. As shown in Figure~\ref{fig:overview}, \projecttitle{} keeps bulk data (values) in the host memory and stores only minimal data (keys + metadata) in the TEE area. The metadata, e.g., hash of the value, timestamps, etc., are kept along with keys in the TEE for integrity verification.

Our work shows how to leverage modern hardware to build efficient, robust, and easily adaptable distributed protocols by meeting the aforementioned transformation requirements.
To achieve our goal, we need to address the following technical questions discussed in~$\S$~\ref{sec:motivation}.
Next, we present the implementation details or our work focusing on four core components of \projecttitle{} ($\S$~\ref{sec:recipe_impl_apis}). Table~\ref{tab:api} summarizes the \projecttitle{}'s API for each component. 

\subsection{System Design Challenges}
\label{sec:motivation}

Our work shows how to leverage modern hardware to build efficient, robust, and easily adaptable distributed protocols by meeting the aforementioned transformation requirements.
Specifically, we address the following research questions.

\noindent{\bf{Q1: How to use TEEs efficiently?}} TEEs are not a panacea: due to their architectural limitations (limited trusted memory and slow syscalls' API)~\cite{shieldstore, treaty}, their naive adoption to build practical systems does not necessarily translate into performance improvements. For example, communication in the state-of-the-art BFT protocols~\cite{10.1145/3492321.3519568, minBFT, 10.1145/3528535.3531516}, which is at the core of any distributed protocol, primarily builds on standardized kernel interfaces (e.g., sockets) suffering from big latencies and not exploiting the full potential of the system~\cite{10.1145/3140659.3080208}. 

In \projecttitle{}, we carefully address these limitations without introducing an additional burden to developers. We build a secure Remote Procedure Call (RPC) framework on top of a direct I/O network stack for TEEs that achieves three goals. First, it boosts performance by avoiding expensive syscalls within TEEs. Secondly, it extends the transferable authentication and non-equivocation primitives across the untrusted network infrastructure, realizing the transformation in practice. Lastly, we follow an established RPC-programming paradigm that has proven to be the most effective one for building distributed protocols~\cite{Hermes:2020, farm, fasst, erpc}.

\noindent{\bf{Q2: How to use TEEs to transform and build \emph{practical} systems?}} 
While Clement et al. show that a translation of a CFT protocol to a BFT protocol {\em exists}, this mechanism adopts an impractical strategy when it comes to building real systems. The entire (transformed) system relies on an expensive mechanism to ensure the correct execution of the underlying CFT protocol. In each round, each replica needs to receive the history of all previous messages, verify their authentication, and replay the execution of the protocol's entire history. This way, it is ensured that non-Byzantine replicas rebuild their state correctly and also that the currently executed message is legitimate (i.e., derives from a valid execution scenario of the protocol).

Secondly, the transformed protocol may amplify the native semantics of the original CFT protocol such as linearizable local reads. As in classical BFT protocols, clients cannot trust individuals, instead, they build collective trust by receiving $f+1$ identical replies from different replicas to ensure that at least one correct replica has responded. 


We design \projecttitle{} to work out-of-the-box to build real systems. \projecttitle{} leverages the properties offered by TEEs to shield the correct protocol execution while our network stack extends the security properties to the network. As a result, our approach does not impose any dependency on the history execution of the protocol, and the original protocol's message complexity is not affected. We also offer an authenticated, per-node, in-memory KV store to allow replicas to detect integrity and authenticity violations and to support local reads individually. Our approach improves performance, but enables easy adoption as well; developers do not have to worry about maintaining protocols' semantics in Byzantine settings.

\noindent{\bf{Q3: How to realize initialization, recovery, and failure detection?}} While the transformation remains agnostic with respect to the transformed CFT protocol in normal operation, the system designers still need to design recovery mechanisms when failures occur. Specifically, Clement et al. do not address how the system \emph{initializes} its state, \emph{detects} failures, and \emph{recovers} from them.  Different CFT protocols have different mechanisms for recovery and failure detection. Some protocols continue to operate when failures occur~\cite{lynch:1997, primary-backup} while others rely on accurate timeouts to detect non-responsive leaders and nodes~\cite{Hermes:2020, raft, chain-replication}. Unfortunately, TEEs come with neither a trusted initialization mechanism for distributed systems~\cite{ias} nor a trusted timer source~\cite{sgxtrustedtime, monotoniccounterssgx}.

\projecttitle{} builds on a secure substrate that overcomes these limitations. We build on a mechanism for collective attestation and a trusted lease mechanism~\cite{t-lease} which is a foundational primitive for \emph{trusted timeouts}, failure detectors~\cite{222603}, leader election~\cite{815321}, etc.

\myparagraph{Q4: Is BFT enough? The case for confidential BFT protocols} Applications that manage sensitive data (e.g., health-care applications~\cite{10.1093/jamia/ocx068}) adopt blockchain solutions for privacy. To this end, cloud-hosted blockchain solutions have been launched~\cite{baasAlibaba, baasAWS, baasAzure, baasIBM, baasOracle}. However, these cloud-hosted blockchain systems that fundamentally build on agreement protocols for serialising the ledger~\cite{rafthyperledger}, jeopardise the blockchain principles of decentralised trust~\cite{10.1145/3528535.3531516}. 

While BFT protocols offer an important foundation to build trustworthy systems, we argue that more and more modern applications~\cite{9484786, 6038579, Chong2007SIFEC, privacy_nlm, cryptdb, trustedDB, ciad} seek confidentiality beyond the scope of the BFT model. \projecttitle{} satisfies this need. Built on top of TEEs, \projecttitle{} transparently offers confidential execution without sacrificing performance.

\subsection{\projecttitle Implementation and APIs}
\label{sec:recipe_impl_apis}
\label{subsec:networkin}
\label{subsec:KV}
\label{subsec:attestation}
\label{non-equivocation-design}

Table~\ref{tab:api} summarizes the \projecttitle{}'s API for each system component.

\myparagraph{\projecttitle{} networking} \projecttitle{} adopts the Remote Procedure Call (RPC) paradigm~\cite{286500} over a generic network library with various transportation layers (Infiniband, RoCE, and DPDK), which is also favorable in the context of TEEs where traditional kernel-based networking is impractical~\cite{kuvaiskii2017sgxbounds}. 

\begin{table}[t]
\small

\begin{center}
\begin{tabular}{ |c|c| }
 \hline
 \bf{Attestation API} &  \\ \hline
 \multirow{1}{*}{\texttt{attest(measurement)}} & Attests the node based on  a measurement.  \\  \hline \hline
 \bf{Initialization API} &  \\ \hline
 \texttt{create\_rpc(app\_ctx)} & Initializes an RPCobj. \\
  \texttt{init\_store()} & Initializes the KV store. \\
  \texttt{reg\_hdlr(\&func)} & Registers request handlers. \\ \hline \hline
 \bf{Network API} &  \\ \hline
 \texttt{send(\&msg\_buf)} & Prepares a req for transmission. \\
 \multirow{1}{*}{\texttt{respond(\&msg\_buf)}} & Prepares a resp for transmission. \\
 \texttt{poll()} & Polls for incoming messages. \\\hline \hline
 \bf{Security API} &  \\ \hline
 \texttt{verify\_msg(\&msg\_buf)} & Verifies the authenticity/integrity and cnt of a msg. \\
 \texttt{shield\_msg(\&msg\_buf)} & Generates a shielded msg. \\ \hline \hline
 \bf{KV Store API} &  \\ \hline
 \texttt{write(key, value)} & Writes a KV to the store. \\
 \hline
 \multirow{2}{*}{\texttt{get(key, \&v$_{TEE}$)}} & Reads the value into \texttt{v$_{TEE}$} \\ & and verifies integrity. \\ \hline 
\end{tabular}
\end{center}
\caption{\projecttitle{} library APIs.} \label{tab:api}
\vspace{-6pt}
\end{table}

\noindent\underline{Initialization.} Prior to the application's execution, developers need to initialize the networking layer by specifying the number of concurrent available connections, the types of the available requests, and by registering the appropriate (custom) request handlers. In \projecttitle{} terms, a communication endpoint corresponds to a per-thread RPC object (RPCobj) with private send/receive queues. All RPCobjs are registered to the same physical port (configurable). Initially, \projecttitle{} creates a handle to the NIC, which is passed to all RPCobjs. Developers need to define the types of RPC requests, each of which might be served by a different request handler. Request handlers are functions written by developers that are registered with the handle prior to the creation of the communication endpoints. Lastly, before executing the application's code, the connections between RPCobjs need to be correctly established.

\noindent\underline{send/receive operations.} We offer asynchronous network operations following the RPC paradigm. For each RPCobj, \projecttitle{} keeps a transmission (TX) and reception (RX) queue, organized as ring buffers. Developers enqueue requests and responses to requests via \projecttitle{}'s specific functions, which place the message in the RPCobj's TX queue. Later, they can call a polling function that flushes the messages in the TX and drains the RX queues of an RPCobj. The function will trigger the sending of all queued messages and process all received requests and responses. Reception of a request triggers the execution of the request handler for that specific type. Reception of a response to a request triggers a cleanup function that releases all resources allocated for the request, e.g., message buffers and rate limiters (for congestion). 

\noindent\underline{API.} We offer a create\_rpc() function that creates a bound-to-the-NIC RPCobj. The function takes the application context, i.e., NIC specification and port, remote IP and port, as an argument, creates a communication endpoint, and establishes a connection with the remote side. The function returns after the connection establishment. RPCobjs offer bidirectional communication between the two sides. Prior to the creation of RPCobj, developers need to specify and register the request types and handlers using the reg\_hdlr() which takes as an argument a reference to the preferred handler function. 

For exchanging network messages, we designed a send() function that takes the session (connection) identifier, the message buffer to be sent, the request type, and the cleanup function as arguments. This function submits a message for transmission. Upon reception of a request, the program control passes to the registered request handler, where the function respond() can submit a response or ACK to that request. Lastly, the function poll() needs to be called regularly to fetch or transmit the network messages in the TX and RX queues.

\myparagraph{Secure runtime} We build our codebase in C++ using \scone{} to access the TEE hardware. \scone{} exposes a modified libc library and combines user-level threading and asynchronous syscalls~\cite{flexsc} to reduce the cost of syscall execution. While we limit the number of syscalls, leveraging  \scone{}'s exit-less approach allows us to optimize the initialization phase that vastly allocates host memory for the network stack and the KV store. To enable NIC's DMA operations and memory mappings to the hugepages (for message buffers and TX/RX queues) ($\S$~\ref{subsec:networkin}), we overwrite the \texttt{mmap()} syscall of \scone{} to bypass its shield layer and allow the allocation of (untrusted) host memory. 

For the cryptographic primitives, we build on OpenSSL~\cite{openssl}. Lastly, we build on a lease mechanism~\cite{t-lease} in \scone{} for auxiliary operations, e.g., failures detection and leader's election.

\myparagraph{\projecttitle{} key-value store} \projecttitle{} provides a lock-free, high-performant KV store based on a skip-list. We partition the keys from the values' space by placing the keys along with metadata (and a pointer to the value in host memory) inside the TEE's memory area, the {\em enclave}, and storing the values in the host memory. Our partitioned KVs reduces the number of calculations for integrity checks, compared to prior work~\cite{shieldstore} which implements (per-bucket) merkle trees and re-calculates the root on each update. Importantly, separating the (keys + metadata) and the values between the enclave and untrusted unlimited memory decreases the Enclave Page Cache (EPC) pressure~\cite{speicher-fast}. 

The developer might want to overwrite/implement the init\_store() function, which will keep an application's state and metadata in the trusted enclave. \projecttitle{} implements its hybrid skiplist based on folly library~\cite{folly}. The write() function updates the KV, while the get() function copies the value of the given key in the protected area. The function also verifies the value's integrity. We implement an allocator for host memory that is given as an initialization parameter to the KV store.

\projecttitle{}'s KV store design resolves Byzantine errors since the metadata (and the code that accesses them) reside in the enclave. That said, \projecttitle{} allows for local reads as nodes can verify the integrity of the stored values. Our partitioned scheme {\em seamlessly} strengthens the system's security properties further and can offer confidentiality by encrypting the values outside the TEE. \projecttitle{}-transformed protocols that further offer confidentiality outperform the BFT systems ($
\S$~\ref{sec:eval}).



\myparagraph{Attestation process} The attestation process is initialized by the \emph{challenger}, a remote process that can verify the authenticity of a specific TEE. The challenger executes the remote\_attestation() function to send an attestation request to the application---usually in the form of a nonce (a random number). The challenger and the application then pass through a Diffie-Hellman key exchange process~\cite{10.1145/359460.359473}. The application generates an ephemeral public key which is used by the challenger later to provision any secrets.

When the TEE receives the nonce, it calls the attest() and generates a \emph{measurement} ($\mu$) of its state and loaded code. Following this, the TEE calls into the generate\_quote$(\mu, k_{pub})$ to sign $\mu$ (quote) with the $key_{hw}$ which is fetched from the TEE's h/w. The TEE signs and encrypts the quote quote$_{\sigma_{k_{pub}}}$ over the challenger's public key $k_{pub}$, which is then sent back to the challenger. Upon successful verifications of the quote$_{\sigma_{k_{pub}}}$, the challenger shares secrets and configurations.

To offer low-latency attestations within the same datacenter that \projecttitle{} runs, we build a Configuration and Attestation service (CAS). The Protocol Designer (PD) deploys the CAS inside a TEE and attests it through the hardware vendor's attestation service---e.g., Intel Attestation Service (IAS~\cite{ias}). Once the CAS is attested, it is trusted, and the PB can upload secrets and configurations. 

The challenger asserts upon a failed verification and denies sharing any secret or configuration data. Otherwise, it distributes all necessary shared secrets.

\section{Evaluation}~\label{sec:eval}
\vspace{-15pt}
\subsection{How to Apply the \projecttitle{} Library?} Developers can use the \projecttitle{}-lib API to transform their preferred CFT protocol for Byzantine settings without further modifying the core states of the protocol. Listing~\ref{lst:transformation} illustrates Raft's transformation using the same example of R-Raft from Figure~\ref{fig:raft_example}.  In Listing~\ref{lst:transformation}, the blue sections show the native Raft code, whereas the orange sections show the modifications introduced by \projecttitle{}.
\vspace{10pt}

\begin{lstlisting}[frame=h,style=customc,
                    label={lst:transformation},
                    caption= Raft transformation using \projecttitle{}: blue sections (native Raft) and orange sections (\projecttitle additions).]
// ----------- Requests handlers definition -----------
void replication_hdlr(Raft_ctx* ctx, Msg* recv_msg) {
    // verifies recv_msg integrity and counter
@\HilightYlinewidth@    [msg, cnt] = verify_msg(recv_msg);
    ... // appends the req to the on-going reqs buffer
@\Hilight@    conn.respond(@\HilightY@shield_msg(ACK_repl)); // transmits ACK
}
void cmt_hdlr(Raft_ctx* ctx, Msg* recv_msg) {
@\HilightYlinewidth@    [msg, cnt] = verify_msg(recv_msg);
    auto [key, val] = decode(req);
    // stores val in host mem and its certificate in TEE
@\Hilight@    ctx->kv.write(key, val);  
@\Hilight@    conn.respond(@\HilightY@shield_msg(ACK_cmt)); // transmits ACK
}
// ----------- Init phase  -----------
auto ctx = new Raft_ctx(metadata, node_type); 
// init local KV with host allocated memory and a cipher
@\Hilight@ ctx->kv = init_store(@\HilightY@HostMemSize, cipher); 
@\Hilight@ RPC_obj conn = create_rpc(@\Hilightsmall@enc_key); // create RPC handle
// registers handlers
@\Hilight@ conn.reg_hdlr(&cmt_hdlr);
@\Hilight@ conn.reg_hdlr(&replication_hdlr);
// ----------- Raft leader -----------
for (auto& node : followers_list) {
    conn.wait_until_connected(node); // connects with followers
}
for (;;) {
    ... // gets client request and marks it as on-going
    for (auto& follower : connections) 
    // generates a shielded message and bcast to followers
@\Hilight@      follower.send(@\HilightY@shield_msg(rep_req), TypeRepl);    
@\Hilight@    conn->poll(); // polls to flush TX/RX queues 
    for (auto& follower : connections) ... // bcast commit
@\Hilight@      follower.send(@\HilightY@shield_msg(cmt_req), TypeCmt);
    ... // commit phase, apply changes to local kv
@\Hilight@    ctx->kv.write(key, val); 
}
\end{lstlisting}

Developers need to port the codebase within the TEE and use the \projecttitle{}'s network API to replace the conventional unsecured RPC-API~\cite{rdma, erpc, f04eb9b864204bab958e72055062748c} with \projecttitle{}-lib's networking functions extending the TEEs' trust across the network. Some of the \projecttitle{}'s API remains equivalent to the native API; typical examples are the poll() and reg\_hdlr() functions. On the other hand, we introduced some slight modifications in send() operation and Initialization APIs.  

\begin{figure}
    \centering
     \includegraphics[width=0.5\textwidth]{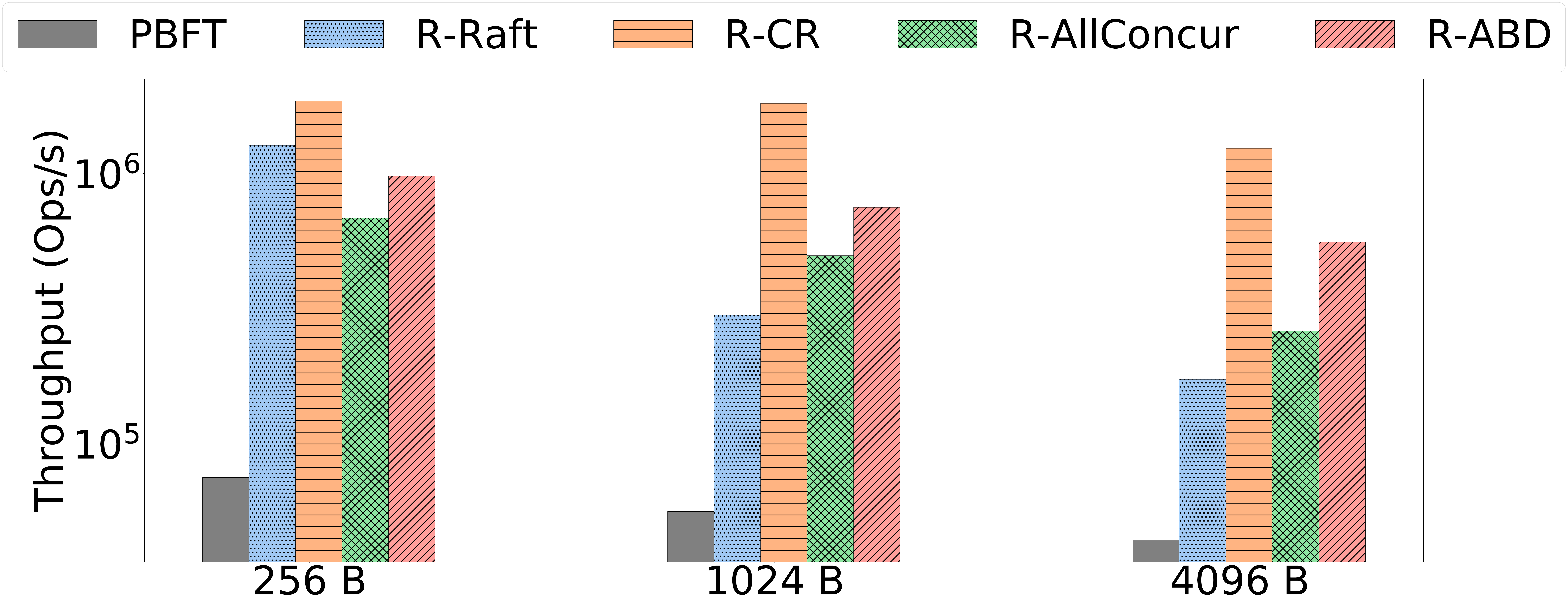}

    \caption{Performance of \projecttitle{} for different value sizes.}
    \vspace{0pt}
    ~\label{fig:value_sz}
\end{figure}

\subsection{\projecttitle{} in Action for CFT Protocols}

\myparagraph{Experimental setup}
We run our experiments in a cluster of three SGX machines (NixOS, 5.15.43) with CPU: Intel(R) Core(TM) i9-9900K each with 8 cores (16 HT), NIC: Intel Corporation Ethernet Controller XL710 for 40GbE QSFP+ (rev 02) and a 40GbE QSFP+ network switch. For the evaluation, we use the YCSB benchmark~\cite{YCSB} (configured with approx. 10K distinct keys with Zipfian distribution) with various R/W ratios and value sizes.

To show the benefits of our approach, we implement four widely adopted CFT protocols (one of each category in $\S$~\ref{sec:background}, Table~\ref{tab:categories}), with the \projecttitle{}-lib API. We build R-ABD, R-Raft, R-AllConcur, and R-CR, which are the \projecttitle{} versions of ABD, Raft, AllConcur, and CR, respectively. We compare these protocols with BFT-smart~\cite{bft-smart}, an optimized version of PBFT~\cite{Castro:2002} and Damysus~\cite{10.1145/3492321.3519568}, the state-of-the-art version of HotStuff~\cite{DBLP:journals/corr/abs-1803-05069} on top of SGX (with $2f$+1). Next, we discuss the characteristics of protocol categories, our chosen protocol, and our evaluation results.

\myparagraph{\underline{A: Leaderless w/ per-key order}}
Protocols in this family agree on a per-key order of writes in a distributed manner. All nodes can coordinate a write that is completed in at least two rounds. A typical example is Classic Paxos (CP) that achieves consensus in three broadcast rounds.  Several works~\cite{10.1145/2517349.2517350, https://doi.org/10.48550/arxiv.2003.11789, phdthesis, Hermes:2020, lynch:1997} simplify the complexity of CP to boost performance. Protocols such as~\cite{10.1145/2517349.2517350, https://doi.org/10.48550/arxiv.2003.11789, phdthesis} can offer consensus in two rounds but fall back to CP if conflicts occur. Others~\cite{Hermes:2020, lynch:1997} execute writes in two rounds, enforcing all messages to be received by all nodes or relaxing the Read-Modify-Write semantics. These protocols offer linearizable reads by executing quorum reads to consult (at least) the majority. Protocols like~\cite{Hermes:2020} where writes need to reach all nodes allow for local reads (at the cost of availability---if a node fails, writes block).

\myparagraph{Choice: ABD~\cite{lynch:1997}} We implemented ABD, a multi-writer, multi-reader protocol with \projecttitle{} (R-ABD). R-ABD offers linearizable (quorum) reads using Lamport timestamps (TS)~\cite{Lamport:1982} for each key-value (KV) pair. R-ABD broadcasts requests to all replicas and waits for acks from the quorum. 

Writes are executed in two rounds of broadcasts. First, the coordinator asks all replicas to hand over the key's TS, which is securely stored inside the TEE (KVs' metadata). Upon receiving a majority of the timestamps, the coordinator creates a higher TS for that key by increasing the highest received TS. Finally, it broadcasts the new KV pair and its new TS to all replicas, which, in turn, insert the KV pair into their KV store. Upon gathering a majority of
acks it replies to the client.


R-ABD (usually) executes reads in one round by collecting all values (and their TS) from the majority. If the majority agrees on the latest seen TS, the coordinator replies to the client. Otherwise, the coordinator chooses the highest TS and invokes the second round of the write-path (for availability).


\begin{table*}[t]
\begin{minipage}[b]{0.4\linewidth}
\centering
            \begin{tabular}{>{\centering\arraybackslash}p{0.18\columnwidth}>{\centering\arraybackslash}p{0.15\columnwidth}>{\centering\arraybackslash}p{0.15\columnwidth}>{\centering\arraybackslash}p{0.15\columnwidth}>{\centering\arraybackslash}p{0.25\columnwidth}}
              \rowcolor{gray!25}
            \textbf{R/W ratio} &  \textbf{R-ABD} & \textbf{R-CR}  & \textbf{R-Raft}  & \textbf{R-AllConcur} \\
            \hline
            $50\%$ & $6.5\times$ & $13.7\times$  & $5.3\times$ & $6.8\times$\\
            \hline
            $75\%$ & $13.3\times$ & $14.8\times$  & $10.05\times$ & $9.4\times$\\
            \hline
            $90\%$ & $13\times$ & $24\times$  & $16.5\times$ & $9\times$\\
            \hline
            $95\%$ & $12.8\times$ & $21\times$  & $10.7\times$ & $9.5\times$\\
            \hline
            $99\%$ & $12.3\times$ & $23\times$  & $9.8\times$ & $10.5\times$\\
            \hline
        \end{tabular}
        \vspace{10pt}

        \label{tab:speedup}
\end{minipage}\hfill
\begin{minipage}[b]{0.5\linewidth}
\centering
\centering
      \vspace{10pt}
       \includegraphics[width=\textwidth]{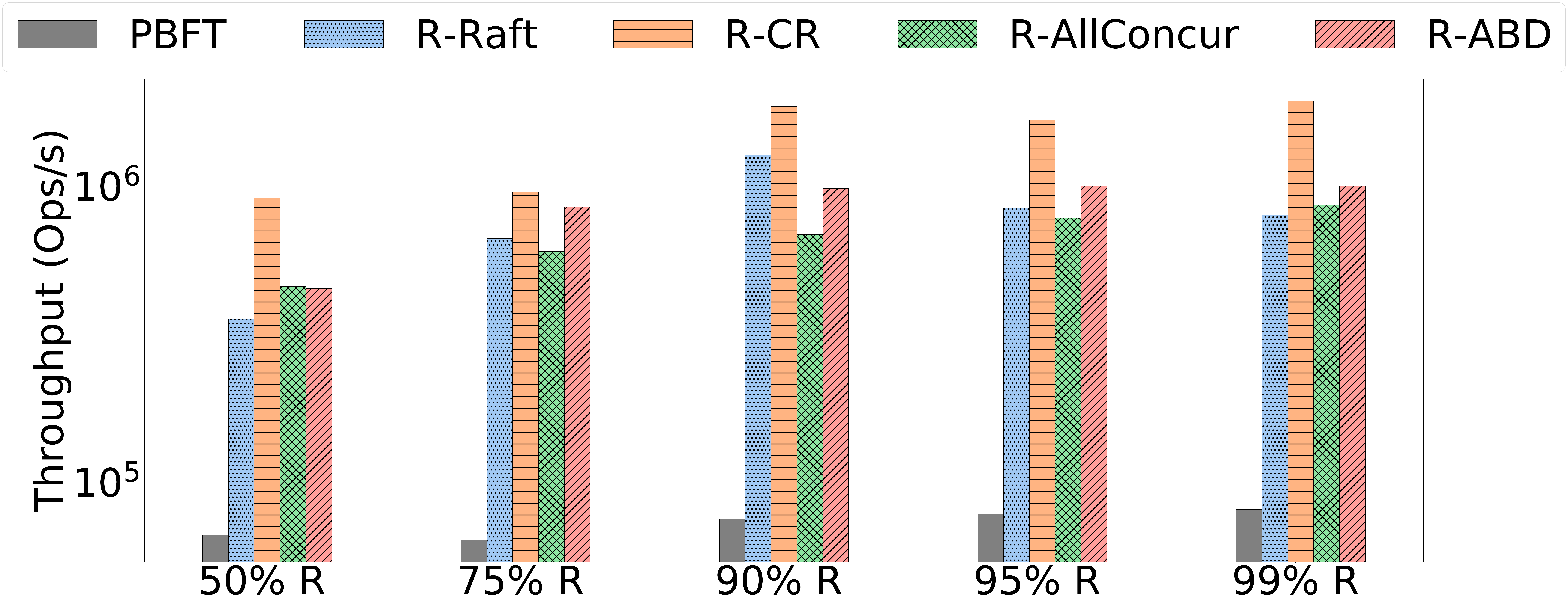}
\end{minipage}
\captionof{figure}{Speedup (left Table) and throughput (right Figure) of four protocols with \projecttitle{} compared with PBFT (BFT-smart).}
\label{fig:end_to_end_perf}
\vspace{10pt}
\end{table*}

\myparagraph{\underline{B: Leader-based w/ total ordering}}
The protocols~\cite{raft, Reed2008AST, 10.1145/2673577} serialize writes at the leader, offering total order. The writes usually require two broadcast rounds; the leader proposes writes to (passive)
followers, which they ack the proposal. Once the leader collects the acks from the majority, the commit round is run, where the nodes apply the proposed writes in their total order.
Since writes are propagated to the majority where the leader is always part of it, the leader can always know the latest committed to write for all keys. As such, leaders can always read locally while followers must forward reads to the leader. 
Some protocols~\cite{Reed2008AST} allow followers to read locally. This is achieved in two ways: they might forego linearizability and downgrade to sequential consistency~\cite{attiya:1991} (with the possibility of reading stale values~\cite{Reed2008AST}), or ensure that all writes reach all followers at the cost of availability.

\myparagraph{Choice: Raft~\cite{raft}} As a representative protocol of this family, we implement Raft with \projecttitle{} (R-Raft). We target linearizability; all reads are forwarded to the leader, which also serializes writes. The leader proposes writing to replicas and commits the request when the majority of them acknowledge the proposal.

The leader stores writes in an uncommitted\_queue inside the TEE. We spawn a dedicated (worker) thread to manage this queue and serialize all writes. The worker thread broadcasts the request (or a batch of consecutive requests) to all followers. The followers verify the messages. As an optimization, followers accept future messages, storing them in a separate queue. The followers commit requests respecting the leader's total order and send acks for one or more consecutive requests. The leader only commits a request and responds to the client when it receives a response from the majority. 

\begin{figure}[t]
\centering

    \includegraphics[width=0.5\textwidth]{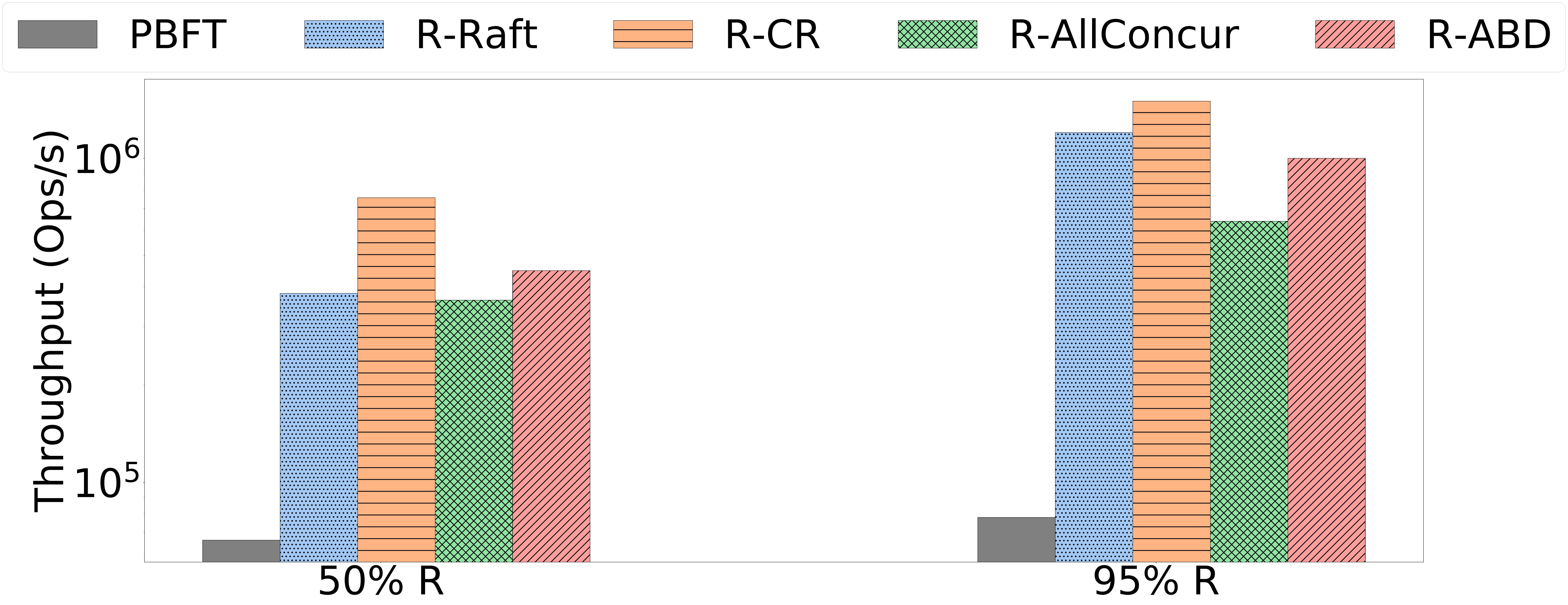}
       
       \caption{Throughput of \projecttitle{} (w/ confidentiality) compared with PBFT (BFT-Smart).}
    \label{fig:confidentiality}
\end{figure}
\myparagraph{\underline{C: Leader-based w/ per-key order}}
Protocols in this class use the leader node to only serialize
writes to the same key. All writes are steered to
the leader node, which ensures that writes to the
same key are applied in the same order by all replicas. 
These protocols can offer linearizability (it is a compositional property), similar to the leader-based protocols with total order. While writes are propagated to a majority of nodes, reads are propagated to the leader. As the protocols do not respect a total ordering, local reads to followers lead to weak guarantees such as eventual consistency~\cite{10.1145/1435417.1435432}. As before, we can allow for local reads to all nodes when writes are guaranteed to propagate to all followers. 

\myparagraph{Choice: Chain Replication~\cite{chain-replication}} As a representative protocol, we implement Chain Replication (R-CR) with \projecttitle{}. In R-CR, the nodes are organized in a chain, through which writes are propagated from the head of the chain to its tail. Similarly to~\cite{f04eb9b864204bab958e72055062748c}, we consider CR among the leader-based protocols as the head node is the leader to serialize all writes.  A write traverses the chain until it reaches the tail where it is considered committed, which guarantees that all writes reach all nodes.  We offer linearizability by reading locally from the tail.  

\myparagraph{\underline{D: Leaderless w/ total ordering}} These protocols rely on a predetermined static allocation of write-ids to nodes. For example, all nodes know that the writes 0 to N-1 will be proposed and coordinated by node-0, the next N writes will be proposed by node-1, and so on. Therefore, in each round each node can calculate the place of each write in the total order based on its own node-id, without synchronizing with any other node. Then, the node broadcasts its writes and their place in the total order. Typically, a commit message is broadcast after gathering acks from a majority of the nodes. Crucially, all nodes must apply the writes in the prescribed total order.

\myparagraph{Choice: AllConcur~\cite{Poke2016AllConcurLC}} To study this category, we implemented AllConcur with \projecttitle{} (R-AllConcur), a decentralized replication protocol with total order that relies on an atomic broadcast primitive. Nodes are organized in a digraph ($G$)~\cite{Poke2016AllConcurLC} where the fault tolerance of the system is given by $G$'s connectivity. For example, to tolerate $1$ node failure on a $3$-node system, we calculated the vertex-connectivity to be equal to $2$; namely, each node is connected to the other two nodes.
For the writes, all nodes track all messages for each round and commit them in a predefined order without synchronization. We can treat reads as writes (for linearizability), or we allow for local reads to replicas offering sequential consistency~\cite{Hunt:2010}.

\subsection{Evaluation Analysis}

\myparagraph{\projecttitle{} vs. PBFT} Figure~\ref{fig:end_to_end_perf} shows the throughput (Ops/s) and the speedup of the four case studies we implemented with \projecttitle{} compared to BFT-smart~\cite{bft-smart} (PBFT) for different read/write workloads (and constant value size/payload, \SI{256}{\byte}). Our evaluation shows that all four protocols with \projecttitle{} outperform the classical BFT $5\times$ to $24\times$. We observe that the local linearizable reads offered by R-CR greatly improve performance. Unfortunately, we see less speedup in read-heavy workloads for the protocols with local reads (e.g., R-Raft and R-AllConcur). We found out that in these protocols, the total ordering was the bottleneck. In the case of R-Raft, the writer thread that serialized all writes was slower than the other worker threads (which executed reads or enqueued writes to the writer thread's queue). Additionally, for R-AllConcur, we saw that collecting all messages for each round decreased throughput. The speedup in R-ABD, R-Raft, and R-AllConcur is moderate for write-heavy workloads where writes require two rounds of messages. R-ABD has a lighter read path; reads require the majority to agree on a value, which is typically resolved in one round. R-CR outperforms R-ABD as reads are done locally. Lastly, as the workload becomes more read-heavy, the throughput is improved slightly due to (1) request rate limiter and (2) single-node bottlenecks.


\begin{figure*}[t]
    \begin{minipage}[t]{.45\textwidth}
        \centering
        \includegraphics[width=\textwidth]{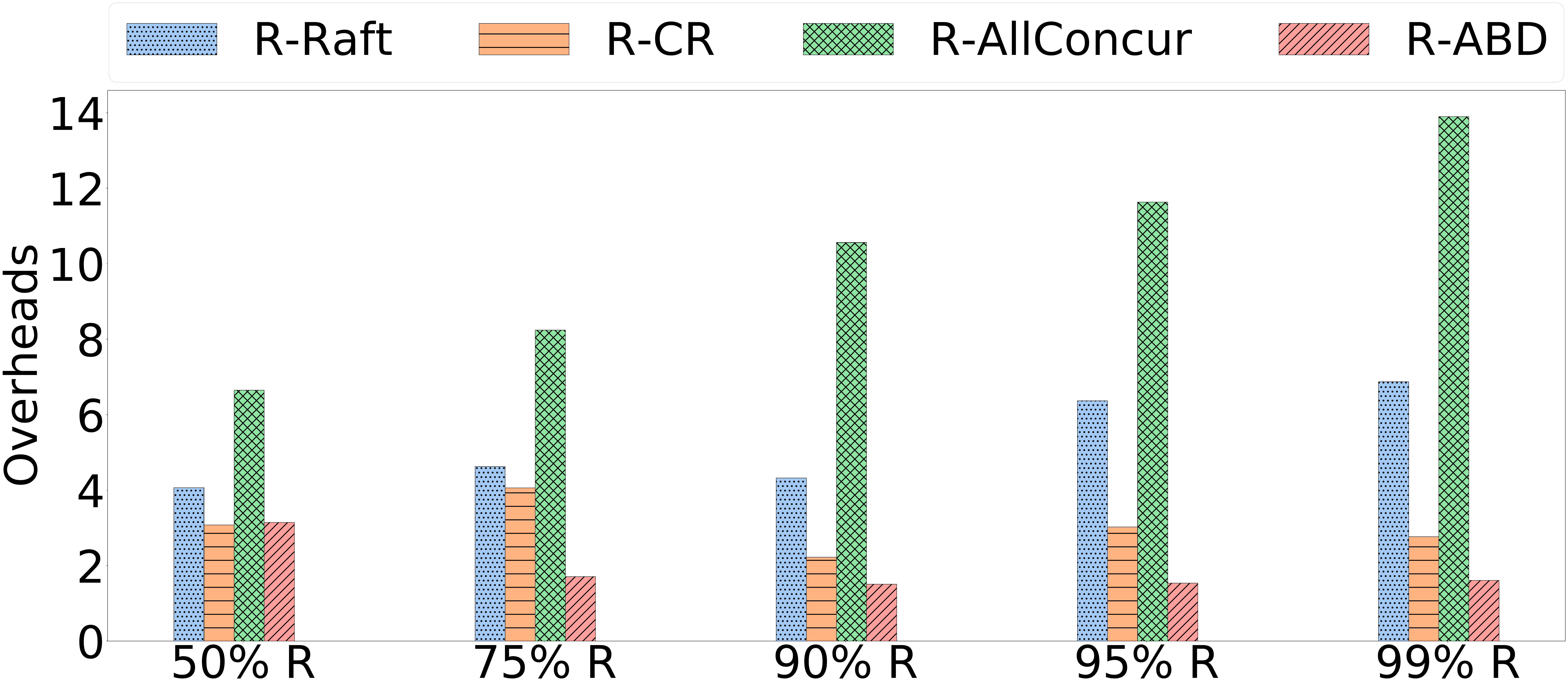}
        \subcaption{Performance overheads of TEEs.}\label{fig:overheads}
        \vspace{2pt}
    \end{minipage}
    \hfill
    \begin{minipage}[t]{.45\textwidth}
        \centering
        \includegraphics[width=\textwidth]{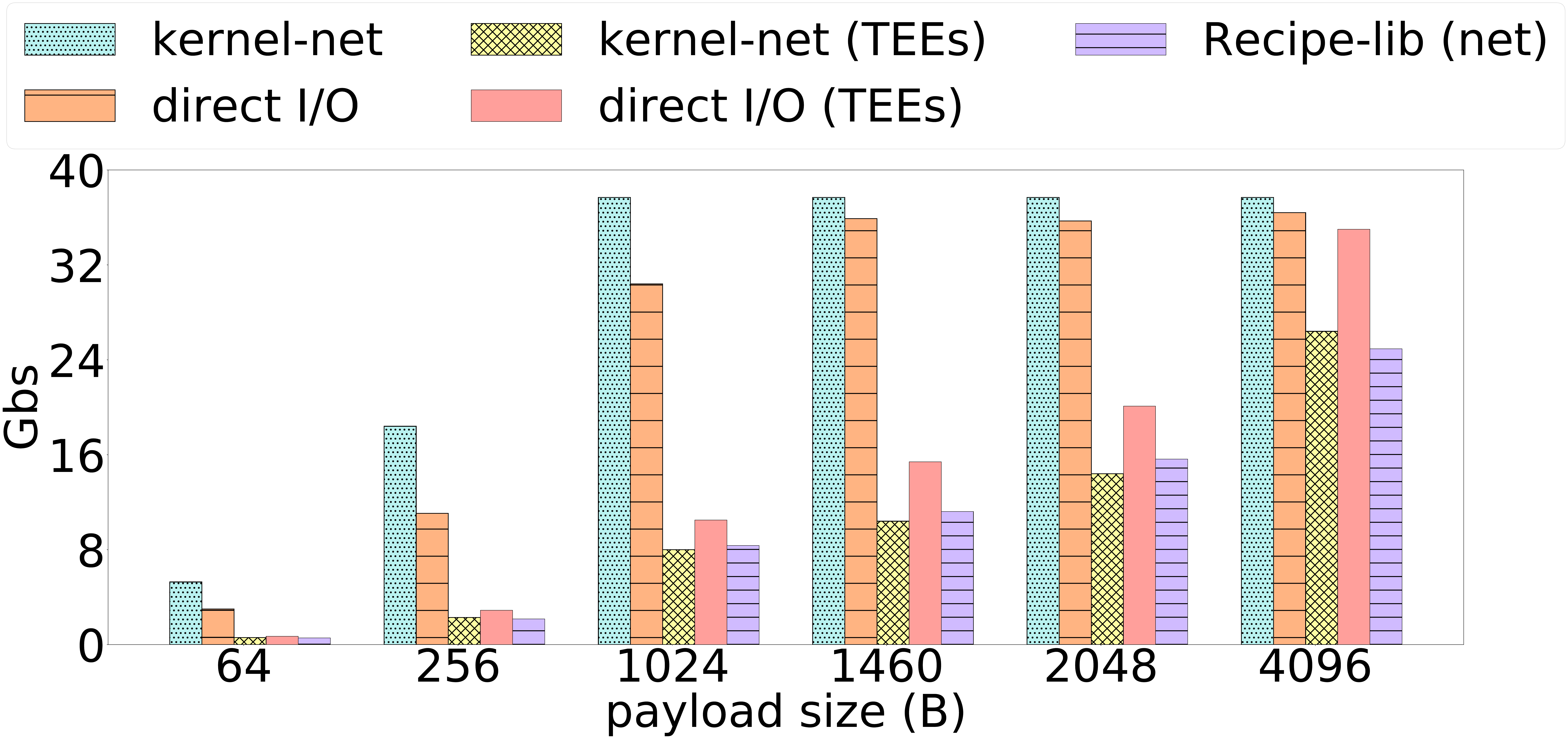}
        \subcaption{Performance evaluation of \projecttitle{}-lib (net).}\label{fig:network}
        \vspace{2pt}
    \end{minipage}  
    \label{fig:1-2}
    \caption{Performance overheads of transformation and TEEs and performance analysis of \projecttitle{} networking.}
    \vspace{5pt}
\end{figure*}

\myparagraph{\projecttitle{} vs Damysus} We compare \projecttitle{} (in h/w) against Damysus~\cite{10.1145/3492321.3519568} with SGX in simulation mode. The setup shows the upper bounds for throughput for Damysus that achieves throughput of $320$ kOp/s, $230$ kOp/s and	$152$ kOp/s for payload sizes \SI{0}{\byte}, \SI{64}{\byte} and \SI{256}{\byte} respectively. Our \projecttitle{} (with \SI{256}{\byte} payload) outperforms $1.1\times$---$2.8\times$ and $2.3\times$---$5.9\times$ Damysus with \SI{0} and \SI{256}{\byte} payloads.

\myparagraph{\projecttitle{} with confidentiality} Figure~\ref{fig:confidentiality} shows the throughput of \projecttitle{} when we also strive for confidentiality; an extra property that is not offered by classical BFT protocols. We guarantee confidentiality by encrypting all data that leave the enclave (network messages, values residing in the host memory). Briefly, the cost for this extra property is a throughput decrement by a factor of 2. Surprisingly, R-ABD shows minimal degradation compared to R-ABD without confidentiality. The reason is that R-ABD quickly saturated all memory resources in our system so the throughput was limited mainly by the requests' rate limiter. We see that even with stronger properties, i.e., confidentiality, \projecttitle{} achieves higher throughput than PBFT: on average we calculate $7\times$ and $13\times$ speedup for $50\%$ and $95\%$ workloads respectively. 

\projecttitle{} with confidentiality boosts throughput up to $4.9\times$ w.r.t. Damysus that does not offer confidentiality.


\myparagraph{Value size} Figure~\ref{fig:value_sz} shows the throughput for different value sizes (under a 90\% R workload) for each of the four protocols. The performance drops as the value size is increased due to the EPC's limited size. While \projecttitle{} places the values and network buffers in the untrusted (unlimited) memory, the bigger the allocations are the more we stress test the (limited) enclave memory. R-Raft and R-AllConcur show the greatest slowdown ($2\times$ to $7\times$ for \SI{4096}{\byte}). We interestingly found out that the batching technique in these protocols with value size of \SI{4096}{\byte} deteriorates the performance and, even, crashes the system by consuming all \scone{}'s memory. For these two protocols with value size \SI{4096}{\byte} we depict the numbers with little ($< 4$) or no batching factor. The other two protocols, R-ABD and R-CR, also show similar behavior. In these protocols we did not use batching as an extra optimization. 

\myparagraph{Transformation and TEEs overheads} Figure~\ref{fig:overheads} shows the overheads introduced by  \projecttitle{} where we compare a native implementation of the protocols with the same network stack without the authentication layer. Overall, an R-CFT protocol experiences $2\times$---$15\times$ slowdown compared to its native execution. The overheads mainly derive from the TEEs. To prove that, we also ran these protocols in \texttt{simulation} mode in \scone{} where the trusted memory (EPC) is unlimited: we found the throughput to be almost equivalent to the native runs' results. Our observation is also explained from the fact that the higher overheads are for AllConcur and Raft. To optimize these protocols we found extremely helpful the batching. However, batching requires allocations/de-allocations of bigger continuous (virtual) memory buffers which stress test \scone{} memory subsystem.

\myparagraph{\projecttitle{}-lib network performance} Figure~\ref{fig:network} shows the network throughput (Gbps) of five competitive network stacks: \emph{(i)} a native and a TEE-based network stack on top of kernel sockets~\cite{iperf}, \emph{(ii)} a native and a TEE-based direct I/O for networking (RDMA/DPDK) and \emph{(iii)} our TEE-based \projecttitle{}-lib network library. This is to isolate the performance gains of the RDMA-based stack in \projecttitle{}.

We deduct two core conclusions. First, TEEs (\scone{}) can degrade network throughput $4\times$---$8\times$ for both kernel-net and direct I/O networking compared to their unprotected (native) runs. Consequently, a naive adoption of TEEs for BFT does not necessarily translate to performance gains. Secondly, \projecttitle{}-lib network performs up to $1.66\times$ faster than the kernel-based networking (kernel-net (TEEs)). As a takeaway the performance speedup ($24\times$ w.r.t. PBFT and $5.9\times$ w.r.t. Damysus) for all our four use-cases with \projecttitle{} are primarily due to the transformation (\projecttitle{}) rather than the use of direct I/O.

\myparagraph{Attestation} Table~\ref{tab:attest} shows the latencies of Intel's Attestation Service (IAS)~\cite{ias} and \projecttitle{} CAS. We found that the (mean) average of our CAS is $0.17$ s, i.e., $~18\times$ faster than the IAS ($2.9$ s).

\begin{table}[t]
\setlength{\tabcolsep}{3pt}
\center
\begin{tabular}{>{\centering\arraybackslash}p{0.225\columnwidth}>{\centering\arraybackslash}p{0.225\columnwidth}>{\centering\arraybackslash}p{0.225\columnwidth}}
  \rowcolor{gray!25}
&  \textbf{Mean \ s} &  \textbf{Speedup}  \\
\hline
\projecttitle CAS & $0.169$ &  $18.2\times$ \\
\hline
IAS & $2.913$  &  \\
\hline
\end{tabular}
\caption{The end-to-end latency comparison between the attestation mechanisms using \projecttitle CAS and IAS.}
\label{tab:attest}
\vspace{-2pt}
\end{table}


\section{Summary of the \projecttitle{} Library}
We present the \projecttitle{} library, a generic library for transforming CFT protocols to tolerate Byzantine failures without any modifications to the core of the protocols, e.g., states, message rounds, and complexity. We realize our \projecttitle{} library by leveraging the advances in modern hardware; we use trusted hardware to guarantee transferable authentication and non-equivocation for thwarting Byzantine errors. Further, we combine trusted hardware with direct network I/O~\cite{rdma, dpdk} for performance. We present an extensive evaluation of \projecttitle{} by applying it to four CFT protocols: Chain Replication, Raft, ABD, and AllConcur. We evaluate these four protocols against the state-of-the-art BFT protocol implementations and show that \projecttitle{} achieves up to $24\times$ and $5.9\times$ better throughput.

\section*{Acknowledgements}
This work was supported in parts by an ERC Starting Grant (ID: 101077577), the Intel Trustworthy Data Center of the Future (TDCoF), and the Chips Joint Undertaking (JU), European Union (EU) HORIZON-JU-IA, under grant agreement No. 101140087 (SMARTY).
The authors acknowledge the financial support by the Federal Ministry of Education and Research of Germany in the programme of "Souverän. Digital. Vernetzt.". Joint project 6G-life, project identification number: 16KISK002. 
%

\balance
\bibliographystyle{plain}
\bibliography{main}

\end{document}